\documentclass{aa}
\usepackage[varg]{txfonts}
\usepackage{graphicx}
\usepackage{natbib}
\usepackage{pifont}
\usepackage{xcolor}
\usepackage{gensymb}
\usepackage{hyperref}
\bibpunct{(}{)}{;}{a}{}{,}

\begin{document}

\title{Impact risk from the circumsolar dust ring on Venus's orbit}

\author{Ariane Courtot \inst{1,2} \thanks{ariane.courtot@obspm.fr} \and Mark Millinger \inst{1}}
\institute{
ESA Space Environments and Effects Section (TEC-EPS), ESTEC, The Netherlands
\and
LTE, Observatoire de Paris, PSL Research University, Sorbonne Universit\'{e}, Université de Lille, LNE, CNRS, France
}
\date{Received 2025 / Accepted date }

\abstract
{A circumsolar dust ring on Venus's orbit was discovered following observations by the Helios spacecraft and then confirmed thanks to observations by STEREO and the Parker Solar Probe. The impact risk it poses needs to be evaluated for any spacecraft crossing the ring.} 
{This study aims to provide a basic model of the dust ring in terms of the distribution of particles (including size distribution) and the density of the ring and orbits and to deduce a first estimation of the impact risk to spacecraft crossing the ring. This impact risk is based on impact flux as well as the direction and speed of impacting particles.} 
{We seek to describe the orbits of dust particles in the ring. We explored two ways to generate initial conditions of particles in the ring: one that involves many assumptions and another that is more reliable. We integrated the second set of orbits for 6000~years and studied their evolution. We then selected particles still inside the ring after this integration and used the orbits to compute the impact risk for a spacecraft chosen as an example, BepiColombo.} 
{We demonstrate that the dust ring will persist over the next 6000~years, extending radially and perpendicularly to the Venus orbital plane because of close encounters with Venus and because of non-gravitational forces. We show that particles tend to accumulate at Venus's orbit, but along its orbit, the variations in density are negligible. We computed the number of particles we expect to find in the ring for radii between 2~$\mu$m and 2~cm (i.e. for masses between $10^{-2}$~kg and $10^{-14}$~kg). The size distribution of these particles is based on the interplanetary dust model IMEM2. Using the particles still inside the ring at the end of the integration, we computed the impact flux on BepiColombo due to the ring and show that it is of a similar level as the interplanetary dust background. Even though the impact flux is not negligible, it is low enough to be considered a very minor threat to spacecraft following an orbit similar to BepiColombo. Finally, we computed the velocity and direction of particles impacting BepiColombo and show that these results are not concerning. } 
{We thus conclude that the dust ring is not a major threat to spacecraft following trajectories similar to BepiColombo. This model should be updated in the future with any new data on the ring, as such data are severely lacking at the moment.} 


\maketitle

\section{Introduction}
Models of dust and meteoroids in the Solar System have been developed to properly compute the impact risk to a satellite  \citep{Poppe_2016,Soja_al_2019,Moorhead_al_2020,Pokorny_al_2024}. However, as exploration of our Solar System expands, new structures are being discovered and have to be taken into account. Dust rings are such structures. For example, a circumsolar dust ring close to Mercury's orbit was discovered \citep{Stenborg_al_2018,Pokorny_al_2023} as well as a dust ring near Earth's orbit \citep{Jackson_Zook_1989,Dermott_al_1994,Reach_2010}.
 
In 2007, an analysis of data from the Helios spacecraft concluded that a circumsolar dust ring exists on Venus's orbit \citep{Leinert_Moster_2007}. This information was confirmed thanks to data from other spacecraft: STEREO and the Parker Solar Probe (\citealt{Pokorny_Kuchner_2019,Stenborg_al_2021}). These spacecraft obtained data about the dust ring because their trajectory allowed them to observe it, but doing so was not in the initial science plan of their missions.

This area of space is crossed every time a satellite aims for Venus, Mercury, or the Sun or when a satellite simply uses Venus in a gravity assist maneuver. For example, Pioneer and Magellan were spacecraft orbiting Venus that had to cross this ring during their travel. They both experienced a significant power loss from the solar arrays during their flight, which was interpreted either as effects from Venus's atmosphere (Pioneer) or simply as a reduction of light reaching the solar arrays (Magellan). However, an alternative explanation could be a degradation of the solar arrays because of impacts. While we are not claiming this is what happened to Pioneer and/or Magellan, it might be a problem in the future.

The Parker Solar Probe also crossed the ring, and analyses have been made of the impacts it felt \citep{Szalay_al_2020,Szalay_al_2021,Malaspina_al_2023}. No spike of impacts have been detected that could be linked with the crossing of the Venus dust ring. This could happen if the dust ring does not produce a high impact flux or if the orbit of the Parker Solar Probe did not take the spacecraft to the densest parts of the ring. The detection could also have failed for some particles depending on sizes and speeds. To properly answer the question of impact risks, one cannot exclusively rely on one spacecraft crossing the ring, and it is thus necessary to model the ring.

More recently, BepiColombo used Venus in a gravity assist maneuver to reach its final destination, Mercury. This was also the case of Juice, in August 2025, while the future ESA mission EnVision, scheduled for 2031-2033, will study Venus. Therefore, being able to assess the risks such a dust ring poses is of the utmost importance.

The existence of the circumsolar Venus dust ring has been confirmed by several papers (see Sect.~\ref{sec:data}). They also give information about its width and height as well as its density. However, this information is not enough to conclude the impact risk that this dust ring poses to satellites that would cross it. To properly evaluate the impact risk from the ring, we need to know the distribution of the particles in the ring as well as their size and mass. We also want to evaluate the velocity and direction of the particles upon impact. To obtain such information, a model of the orbits of the particles is necessary, which we develop in this study. We also need to find a suitable model of size and mass distribution of the particles in the ring.

With this study, we only seek to answer these questions, and we do not pretend that this work presents the final model to study the ring. Indeed, a lot of data are missing in order to properly model this ring, and we therefore only aim to propose an initial model dedicated to impact risk that can be used as a basis and expanded upon.

In the next section, we detail the observational data as reported in previous papers. Then we explain our method to obtain the model of the dust ring, especially in terms of the orbits of the particles inside the ring. In Sect.~\ref{sec:res}, we show the results we obtained: the distribution of the particles inside the ring, their velocity, and the evolution of the dust ring. We then evaluate the impact risks posed by the dust ring by using BepiColombo as a test case. Finally, we discuss our results in Sect.~\ref{sec:discussion} and conclude in Sect.~\ref{sec:ccl}.

\section{Data}\label{sec:data}

\subsection{Presentation of data}
In 2007, a first study found evidence of a circumsolar dust ring on Venus's orbit in the data from the Helios mission \citep{Leinert_Moster_2007}. This mission consists of two spacecraft that crossed Venus's orbit and were each equipped with three photometers, each pointing in different directions (10, 30, and 90 degrees away from the ecliptic). After an analysis of which photometers were the most reliable (the 90-degrees photometer on board Helios B), Leinert and Moster cleaned up the data and confirmed the existence of a higher density of dust along Venus's orbit, resembling a ring. They claimed this ring is lying just outside the orbit of Venus. They then tried to compute its radial extent and its thickness. They were also interested in computing how much higher the density in the ring is compared to the interplanetary dust. They showed that, in thickness, the ring is symmetrical to the orbit of Venus. They deduced a radial extent of 0.06 to 0.08~au. The thickness of the ring and its overdensity are two linked parameters, which makes their computation difficult. However, they deduced that, for an overdensity of +10\%, the thickness of the ring would be 0.048~au.

The existence of such a ring was not necessarily a surprise since the existence of the dust ring near Earth's orbit was already known at the time. However, it was still necessary to confirm its existence, and to determine its geometry and density. Using data from several missions, other authors were able to participate in this discussion.

\citet{Jones_al_2013} used images from the STEREO mission to confirm the presence of the ring. In a later paper, they added some precisions to their computations \citep{Jones_al_2017}. They selected relevant images where the data was clear enough. According to their work, the dust ring has a peculiar shape, different from the Gaussian of the Earth dust ring. They believed a two-steps model best represent their data. The inner radius of the ring would be 0.71524~au according to their later article. From this radius to a second radius of 0.73917~au, the ring would be following a specific density distribution. From this second radius to an undefined outer edge, the density would follow another distribution. They also mentioned a general overdensity of +10\% in their first paper and of +8.5\% in their second. Finally, they showed that the ring probably does not follow the same inclination and the same longitude of ascending node as Venus's orbit, but noted that the values of these two orbital elements cannot be given with precision. Interestingly these papers seem to be the only ones that consider this two-step approach of the dust ring. They also went against the previous paper by stating that the inner edge of the ring must be inside the orbit of Venus (and the outer edge outside of Venus's orbit). No values are given for the thickness and the radial extent of the ring.

In 2019, a first dynamical model was developed by \citet{Pokorny_Kuchner_2019} in order to study the origin of the dust ring. The expectation was that it would be made of interplanetary dust being trapped in resonance with Venus. However they showed that dust trapped in resonance is not sufficient to explain the ring. Thus they proposed that the dust would come from hypothetical co-orbital asteroids. They claimed that these asteroids could exist even if they have not be detected until now, because of their proximity to the Sun as seen from Earth. Their preferred model included dust from these co-orbital asteroids, but also from Jupiter-family comets. In their model, they reproduced the data from STEREO and Helios when using a +9.5\% overdensity and a radial extent of 0.06~au. This last figure is independent of size distribution, they claimed. Finally, they also showed that particles smaller than 60~$\mu$m were not dynamically stable and would thus escape from the ring and blend in with the zodiacal cloud very quickly.

Another study on the ring was performed in 2021 \citep{Stenborg_al_2021}, using this time data from the Parker Solar Probe and its WISPR instruments. The specificity of this spacecraft is its high velocity, which requires a specific technique to access the data from WISPRs. They applied to the relevant images this technique, that they had tested previously, which allowed them to confirm the existence of a ring with a thickness of 0.043~au (with a precision of 0.004~au) and an overdensity between +8.6\% and +10.8\%, both values in line with previous papers. They also computed the radial width of the ring along the line of sight and propose a value of 0.093~au, or even 0.108~au. These values are much larger than others reported here, and the authors were very cautious about this value. Their conclusion only reported on their evaluation of the perihelion of the ring (between 0.28 and 0.5~au) and its aphelia (between 0.72 and 1.0~au).

We note that \citet{Stenborg_al_2021} claimed \citet{Jones_al_2013,Jones_al_2017} proposed 0.055~au as the thickness for the ring. However, this interpretation seems quite strong. Indeed, Jones and colleagues proposed the following formula for the density of the ring: $n(r,z) = n(r) exp(\frac{-|z|}{\sigma_z})$, with $r$ as the distance to the Sun and $z$ as the distance to the ecliptic plane. Significantly, the value of $\sigma_z$ could be 0.05478~au \citep{Jones_al_2013} or 0.0518~au \citep{Jones_al_2017}. However, it does not seem to translate to a thickness close to 0.055~au.

\subsection{Choice of data}
From these discussions, we chose the following data as characteristic of the dust ring. Those values might get updated by future missions, but it would not alter the method outlined in the following section. First, all papers proposed an overdensity with respect to classical interplanetary dust model close to 10\%, although usually smaller. We chose this value, as it allowed us to compute the maximum risk for a spacecraft crossing the dust ring. As only \citet{Jones_al_2013,Jones_al_2017} made a case for the two-steps density model, we did not take it into account.

About the radial extent of the ring, only \citet{Leinert_Moster_2007} and \citet{Pokorny_Kuchner_2019} tentatively agreed on a value: 0.06~au is considered an acceptable value in both cases. Interestingly, \citet{Stenborg_al_2021}, that made a very different approximation of the radial extent of the ring (0.093 - 0.108~au), used observations of the ring that are not as reliable. This is because of the geometry of the observations. Finally, \citet{Jones_al_2013,Jones_al_2017} seemed to consider a narrower ring, although a precise value is not given (they defined an inner radius and a middle radius, but not an outer radius). We use $w$ to refer to this radial extent.

About the thickness of the ring, \citet{Leinert_Moster_2007} computed a value of 0.048~au, which was almost reached by \citet[0.047~au is the upper limit given]{Stenborg_al_2021}. \citet{Pokorny_Kuchner_2019} did not give a value and all signs from \citet{Jones_al_2013,Jones_al_2017} point to a larger ring in thickness (the density according to their model should reach negligible values for $|z| > 0.055$~au). Therefore we chose the value of 0.048~au in thickness, which seems an acceptable compromise.

While \citet{Leinert_Moster_2007} claimed the ring lies outside the orbit of Venus, other studies did not confirm this. \citet{Jones_al_2013,Jones_al_2017} described a ring with an inner edge inside Venus's orbit and an outer edge outside Venus's orbit. They also claimed that the maximum density of the ring is outside Venus's orbit, presumably implying the ring should not be centred on Venus's orbit. \citet{Stenborg_al_2021} did not comment on this, only using the term `near' to describe the position of the ring with respect to Venus's orbit. This is because the geometric configuration of the spacecraft's observations does not allow for information on this. Finally, in \citet{Pokorny_Kuchner_2019} model, the ring is centred on Venus's orbit and co-orbiting with the planet. Thus, we chose to centre the ring on Venus's orbit for two reasons. First it is a convenient way to model the ring, that should not have a major impact on our model. Second, it is the only choice that is both fully consistent with \citet{Pokorny_Kuchner_2019} and is acceptable with \citet{Jones_al_2013,Jones_al_2017}.

\citet{Leinert_Moster_2007} claimed that the ring has to be symmetrical with respect to Venus's orbital plane, and most of the other papers did not contradict this statement. Therefore, we chose to follow this indication as well.

In the end, our model proposes orbits corresponding to a dust ring of width $w$ = 0.06~au and of thickness 0.048~au, centred on Venus's orbit, both radially and latitudinally. Its density is +10\% compared to classical models of the surrounding interplanetary dust cloud. Any azimuthal variation is not modelled, as there is not enough data on them. No hypothesis is proposed on the variation of number density since there is also a lack of data on this issue.

\section{Method}\label{sec:meth}
To compute the impact risks, we first tried to investigate orbits that could be representative of the dust ring. We defined a first set of initial conditions (IC1) with several simplifying assumptions. We integrated the particles from this first model, and we used the results to define a second set of initial conditions (IC2) that relies on fewer assumptions. Then we performed a second integration. We then filtered the orbits resulting from this second integration, selecting only particles belonging to the ring. They are the basis for the impact risks computations. As \citet{Pokorny_Kuchner_2019} have shown, orbits in the ring tend to be stable over long timescales so selecting orbits belonging to the ring after our integration of 6000~years makes the impact risks computation more reliable.

The integrations themselves are performed using a RADAU integrator \citep{Everhart_1985}. The RADAU is characterised in part by its automated computation of the length of each time increment. We added two non-gravitational forces (NGFs) to the integrator: Poynting-Robertson drag and solar radiation pressure \citep[for a precise description of the equations, see][]{Olsson-Steel_1987}. The bulk density of the particles was chosen to be 1000~kg/m$^3$. To properly compute the gravitational forces from all eight planets, the Moon, and the Sun, we used the ephemeris INPOP \citep{Fienga_al_2009}. We could have added the Lorentz force, but previous works on meteoroids without this force has shown us that it is not necessary to obtain a clear picture of dynamical evolution \citep{Vaubaillon_al_2005,Courtot_al_2023}. A study has already been done on the effect of the Lorentz force on the dust ring \citep{Zhou_al_2021}.

We integrated the particles from the first set of initial conditions for 2000~years and from the second set for 6000~years. In the case of IC1, we do not seek to model the evolution of the ring during its whole lifetime, but only to estimate the trends in its evolution. Since the goal of this first set of initial conditions is to be a basis for the choices of the second, there is no need to evaluate the results of a full integration. We therefore had to find a compromise between a long integration time, demanding in terms of computational resources, that would describe the evolution of the ring in too much details, and a small integration time that may leads us to miss some important trends. 2000~years seemed a good value in this context. For IC2, we did want to obtain a general view of the evolution of the ring. We are aware that even 6000~years are not enough to completely model the entire evolution, but it is enough to obtain a good view of the evolution of the ring, to assess the reliability of our choices of initial conditions and to be able to select particles to compute the impact risk, the goal of this paper. We come back to these choices in Sect.~\ref{sec:discussion}.

Our code detects close encounters with a planet. We define a close encounter as the moment when the particle is closer to the planet than its Hill radius. We also detect when a particle crashes in a planet, which happens when the distance between the particle and the centre of the planet is smaller than the radius of the planet. When that happens, the particle is removed from the integration. Finally, two tests were done throughout the integration: whether the particle's heliocentric distance is smaller than 0.02~au, which would mean that the particle would spiral towards the Sun quite quickly, or whether it is higher than 1000~au, which would mean that the particle is ejected from the Solar System. These two conditions were never fulfilled because of the integration time. This is not a problem, as we only study the dust particles near Venus's orbit.

Table~\ref{tab:orb_elts} gives the orbital elements for both IC1 and IC2. The reasons for these choices are given below.

\begin{table*}
    \caption{Range of each orbital element of each set of initial conditions (the mean anomaly varies between 0 and $360\degree$ in both cases).}
    \label{tab:orb_elts}
    \centering
    \begin{tabular}{c c c c c c}
    \hline \hline
          & a~(au) & e & i~($\degree$) & $\omega$~($\degree$) & $\Omega$~($\degree$) \\ 
         \hline
         IC1 & 0.693 - 0.753 & 0 - 0.048 & 3.395 & 131.53 & 76.68 \\
         IC2 & 0.633 - 0.813 & 0 - 0.072 & 1.004 - 5.786 & 11.23 - 130.0 & 0.0 - 360.0 \\
         \hline
    \end{tabular}
\end{table*}

\subsection{Simplified set of initial conditions}\label{subsec:first_model}
For the first set of initial conditions IC1, we focused on the variations in semi-major axis $a$ and eccentricity $e$. Initially, the other orbital elements are fixed. The chosen values for each of them is equal to the values of Venus's orbit. We thus have the inclination $i = 3.395 \degree$, the longitude of the ascending node $\Omega = 76.68 \degree$, and the argument of perihelion $\omega = 131.53 \degree$.\footnote{NASA, Planetary Fact Sheet, visited on May 2025. \url{http://nssdc.gsfc.nasa.gov/planetary/factsheet/}} This hypothesis is a very strong one, especially since it is directly contradicted by \citet{Jones_al_2013,Jones_al_2017}. However, as a second step, we did add a variation on the Z-component of the position of the particles that modifies the three angles ($i$, $\Omega$, $\omega$). Furthermore, IC1 is only a basis for IC2, which is much more generic. Therefore all the assumptions detailed for this first set of initial conditions, all very strong, should not be taken as underlying any of the results given in Sect.~\ref{sec:res} and \ref{sec:density_risk}.

The semi-major axis follows this rule, with $a_V = 0.723$~au as the semi-major axis of Venus:
\begin{equation}
    a_V - \frac{w}{2} < a < a_V + \frac{w}{2}.
\end{equation}
This gives us the minimum and maximum value, namely, 0.693~au and 0.753~au.

\begin{figure}
    \centering
    \includegraphics[scale = 0.35]{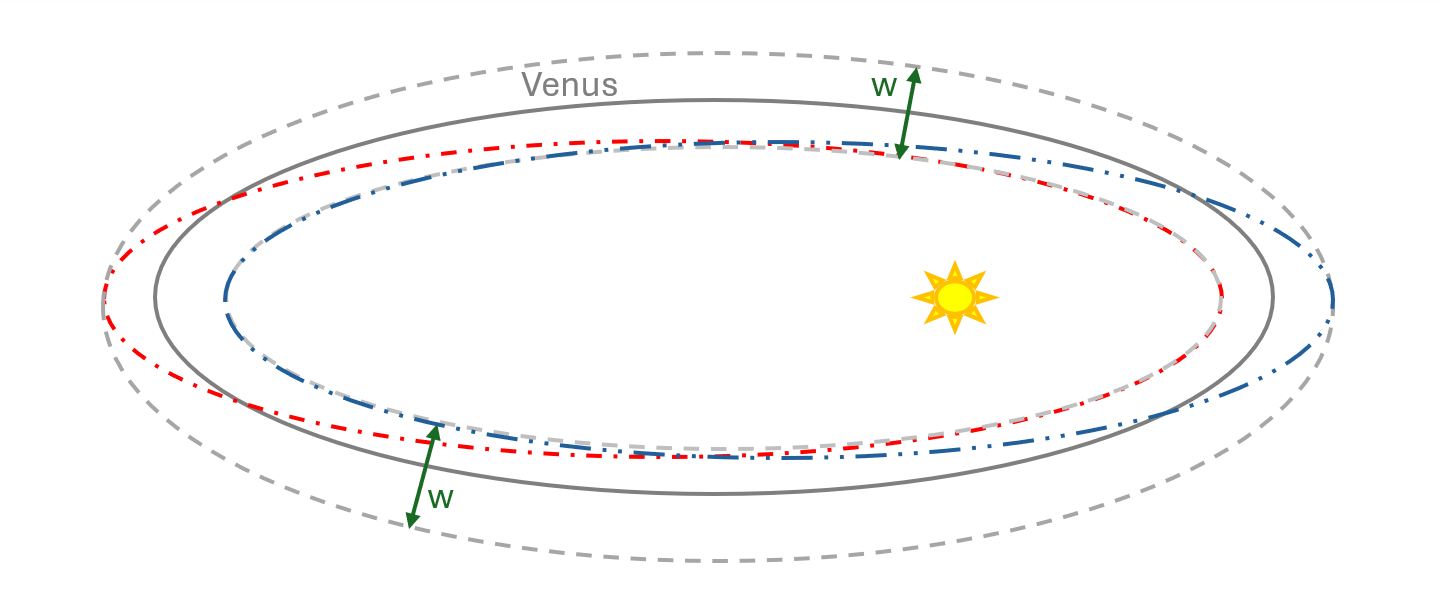}
    \caption{Schematics describing the orbits taken into account for the first model of the dust ring. The dashed grey lines show the outline of the dust ring on the interior and the exterior. The red (dash dot) and blue (dash dot dot) lines represent the two orbits chosen to have maximum and minimum values of eccentricity. All eccentricities here are highly exaggerated for ease of representation.}
    \label{fig:schema}
\end{figure}

To choose the maximum and minimum values of the eccentricity, we noted that the perihelion and aphelion of each orbit must be inside the dust ring. Figure~\ref{fig:schema} shows the orbits used to compute the eccentricities. The grey continuous line is the orbit of Venus, with a highly exaggerated eccentricity for easier visualisation. The dashed grey lines represents the outer and inner outline of the dust ring. We also defined two orbits (the dash-dot red line and dash-dot-dot blue line in the figure) that we assumed are close to the maximum (red orbit) and minimum (blue orbit) eccentricities available in the ring if we only allow the semi-major axis and the eccentricity to vary. 

To compute this maximum eccentricity, we wrote 
\begin{equation}
    \begin{cases}
        Q_{red} = Q_{V} + \frac{w}{2} \\
        Q_{blue} = Q_{V} - \frac{w}{2}
    \end{cases}
    ,
\end{equation}
with $Q$ as the aphelion distance, `red' and `blue' referring to the orbits defined in the previous paragraph, and $V$ referring to Venus's orbit. Similar equations can be written for the perihelion, but they lead to the same results. From these equations, we obtained\begin{equation}
\begin{cases}
    e_{red} = e_V + \frac{w}{2a_V}\\
    e_{blue} = e_V + \frac{w}{2a_V}
\end{cases}
.
\end{equation}

This leads to a maximum eccentricity, $e_{red}$, of 0.048. The equation for the minimum eccentricity, $e_{blue}$, leads to a paradox: a negative eccentricity. This is because in order for the blue orbit to exist, we should have $2a_Ve_V > w$, which is not possible. The only reason why it appears possible in Fig.~\ref{fig:schema} is the exaggerated eccentricity of Venus's orbit, which we chose for better visualisation.

To get the minimum eccentricity, we verified that an orbit with an eccentricity equal to zero could exist. To check this, we verified that a semi-major axis $a_0$ exists such that
\begin{equation}
    \begin{cases}
        Q_V + \frac{w}{2} > Q_0 > Q_V - \frac{w}{2} \\
        q_V + \frac{w}{2} > q_0 > q_V - \frac{w}{2} \\
        b_V + \frac{w}{2} > b_0 > b_V - \frac{w}{2}
    \end{cases}
    ,
\end{equation}
with $q$ as the perihelion distance, $b$ as the semi-minor axis, and all quantities with subscript zero referring to an orbit with eccentricity nul and all other orbital elements chosen equal to Venus. Values of $a_0$ exist that would fulfil the conditions described above, which tells us that a circular orbit should be taken into account in the model. Thus the eccentricity varies between 0 and 0.048. These choices of the semi-major axis and the eccentricity ensure that both the width of the ring and the fact it is centred on Venus's orbit are properly modelled. 

To populate our model, we randomly chose the semi-major axis and the eccentricity of the particles following a uniform distribution inside the bounds computed above. Using these values, a randomly chosen mean anomaly (also uniform distribution), and the values of Venus's orbit for the other orbital elements, we created 33~333 particles, all with a radius of 10~mm. 

Once these particles were created, we computed the state vector of each of them and added a random variation in height, making sure that the Z component of the state vector respected the data on the height of the ring. This allowed us to model the height of the ring as well as its symmetry with respect to Venus's orbital plane. At this point, $i$, $\Omega$, and $\omega$ are no longer fixed to Venus's values. We then repeated this process two more times with a different radius (1~mm, then 0.1~mm). Thus we obtained a model of 99~999 particles with three different radii possible and that follows the input assumptions. The three radii chosen are typically relevant to spacecraft impact risk assessment, with 0.1~mm having an effect on surfaces, 1~mm creating component failure, and 1~cm possibly breaking up the spacecraft.

This method to build IC1 is not very reliable, especially regarding the angles $i$, $\Omega$, and $\omega$. Even the computation of the maximum and minimum eccentricities could be improved, but the goal of IC1 was to be used as a baseline on which to expand. Indeed, as we integrated this ring, we left all orbital elements free to evolve in time. From the intervals of variation for each orbital element, we constructed a second set of initial conditions.

\begin{figure}
    \centering
    \includegraphics[scale = 0.6]{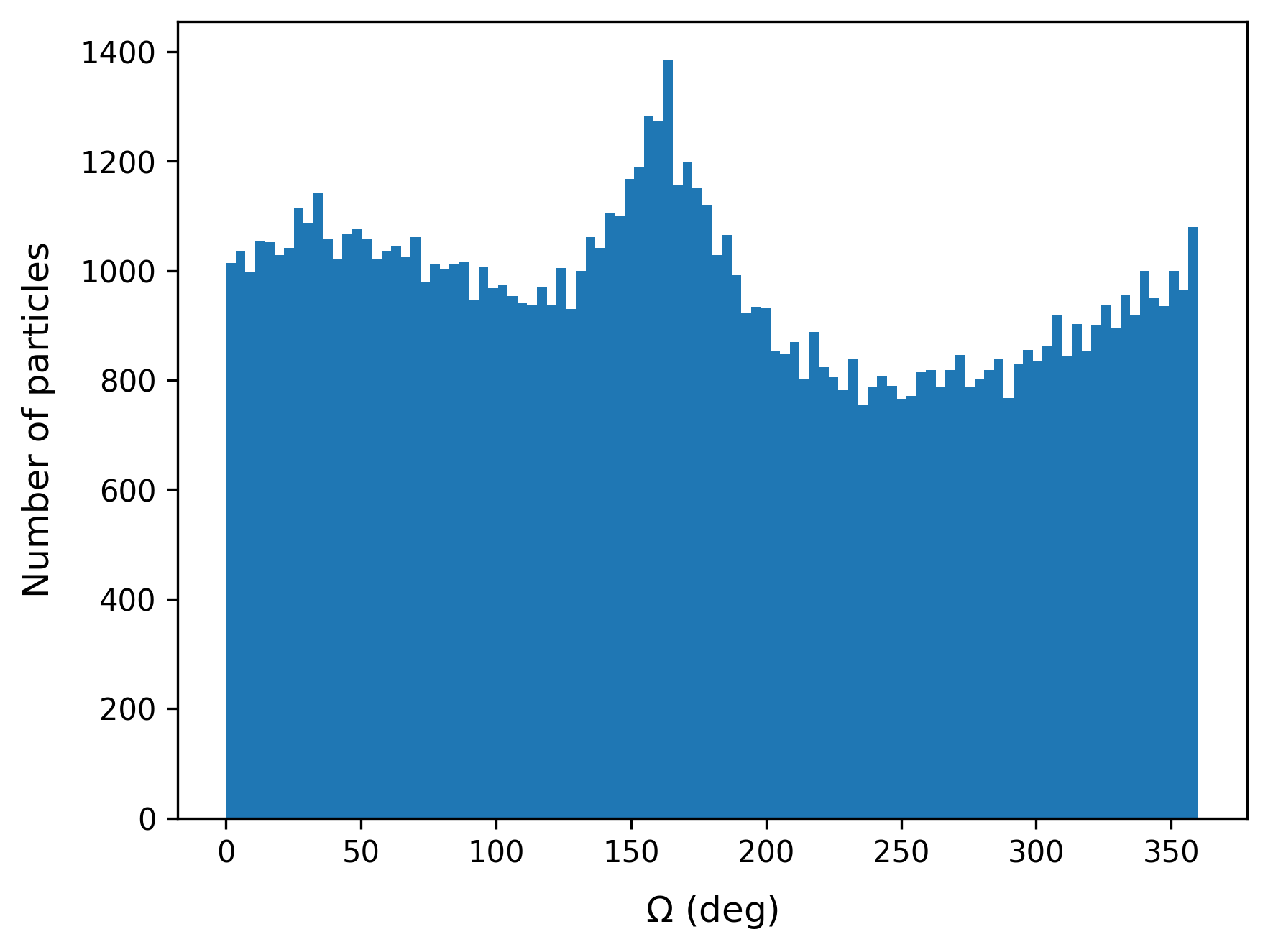}
    \includegraphics[scale = 0.6]{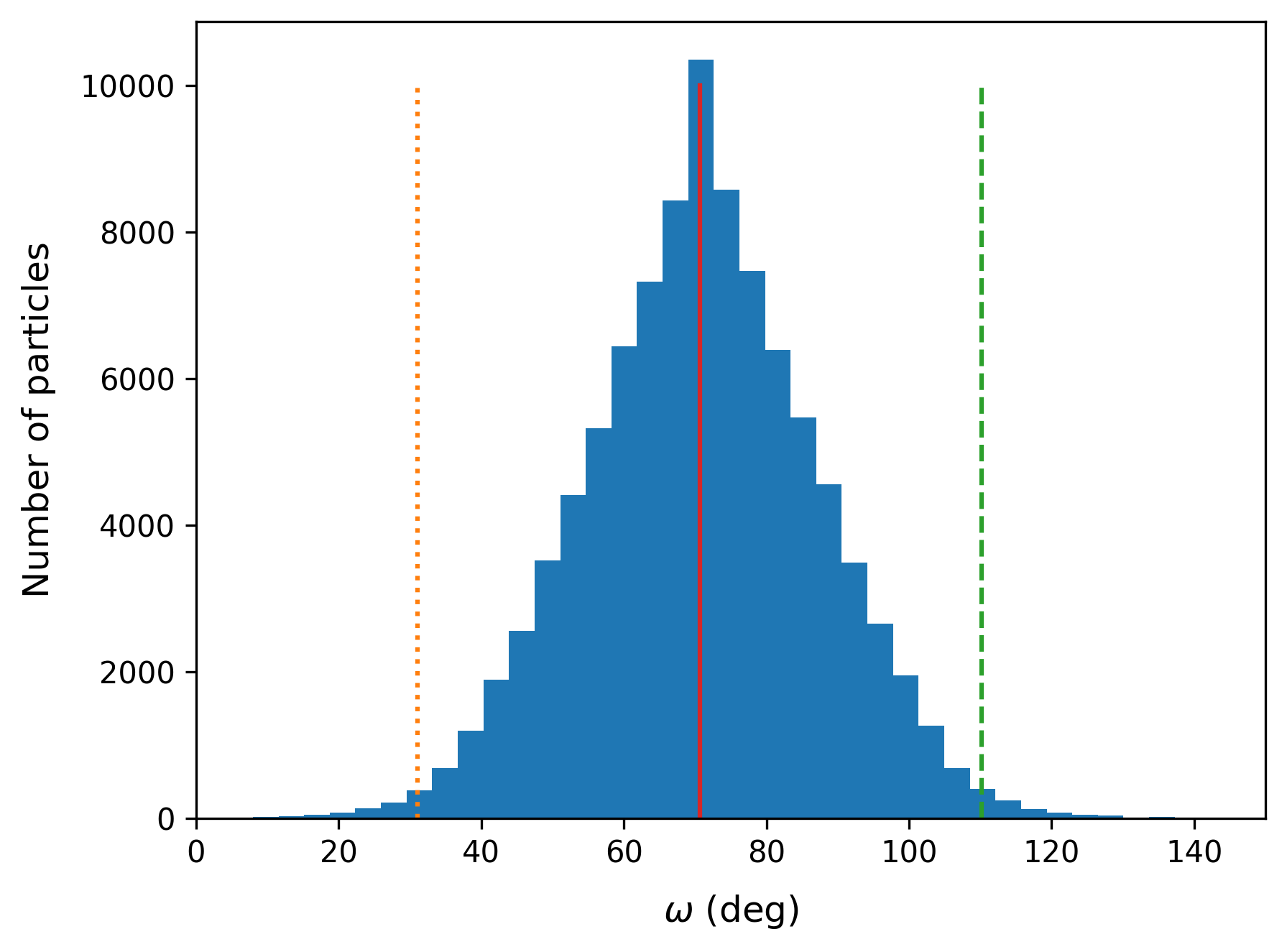}
    \caption{Histograms of the distribution of the longitude of ascending node ($\Omega$) and of the distribution of the argument of perihelion ($\omega$) after the end of the integration (2000~years) of IC1. For the distribution of the argument of perihelion, the continuous red line in the middle is the value of the mean, the dotted orange line on the left corresponds to the value of the mean minus three standard deviation, and the dashed green line on the right shows the value of the mean plus three standard deviation.}
    \label{fig:choice_omegas}
\end{figure}

At the end of the integration, we observed the distribution of orbital elements of the particles final states. This was done to build IC2. The variation of the longitude of ascending node and the argument of perihelion from IC1 were used to constrain the choices of these elements for IC2. Figure \ref{fig:choice_omegas} shows the distribution of those two quantities from the first set of initial conditions at the end of the integration.

\subsection{A more reliable set of initial conditions}\label{subsec:second_model}

To build IC2, we chose orbital elements randomly (following a uniform distribution) in intervals selected according to the results from the IC1. We also added one condition: Each particle has to stay in the ring for at least one period. The point of this condition is to slightly restrict the particles generated, as we want to avoid studying particles that obviously do not belong to the ring. This does not prevent us completely from taking into account particles that only cross the ring without staying in, but it should lessen the effect. 

To check this condition, we verified that, for each particle and each point $P$ in that particle orbit, there exists a point on Venus's orbit that distance to $P$ is compatible with the definition of the ring. Mathematically, for each particle this is written as follows:
\begin{equation}\label{eq}
    \forall M, \exists M_V, 
    \begin{cases}
        |Z_{V}(M_V) - Z(M)| < \frac{H}{2}\\
        \sqrt{(X_V(M_V) - X(M))^2 + (Y_V(M_V) - Y(M))^2} < \frac{w}{2}
    \end{cases}
    ,
\end{equation}
with $M$ the mean anomaly of the particle, $X$, $Y$, and $Z$ describing the position of the particle in a heliocentric reference frame, $H$ the height of the ring and the subscript $V$ denoting the same definitions for Venus. 

The intervals chosen for each orbital element were all informed by IC1. The semi-major axis and the eccentricity intervals chosen in the simplified set both seem reliable, but we were concerned that some values were not taken into account that would in fact be part of the ring. Thus we added a margin of 50\% in the following way: the semi-major axis interval is between $a_V + 1.5\frac{w}{2}$ and $a_V-1.5\frac{w}{2}$. The minimum eccentricity is still zero, but the maximum is now 50\% higher than previously.

The longitude of the ascending node and the argument of perihelion intervals were both chosen based only on the results from IC1. More precisely, the histograms presented in Fig.~\ref{fig:choice_omegas} show that the longitude of the ascending node varies between 0 and $360\degree$. As for the argument of perihelion, the interval is between the mean minus three standard deviation ($11.23\degree$) and the mean plus three standard deviation ($130\degree$). 

\begin{figure}
    \centering
    \includegraphics[scale = 0.12]{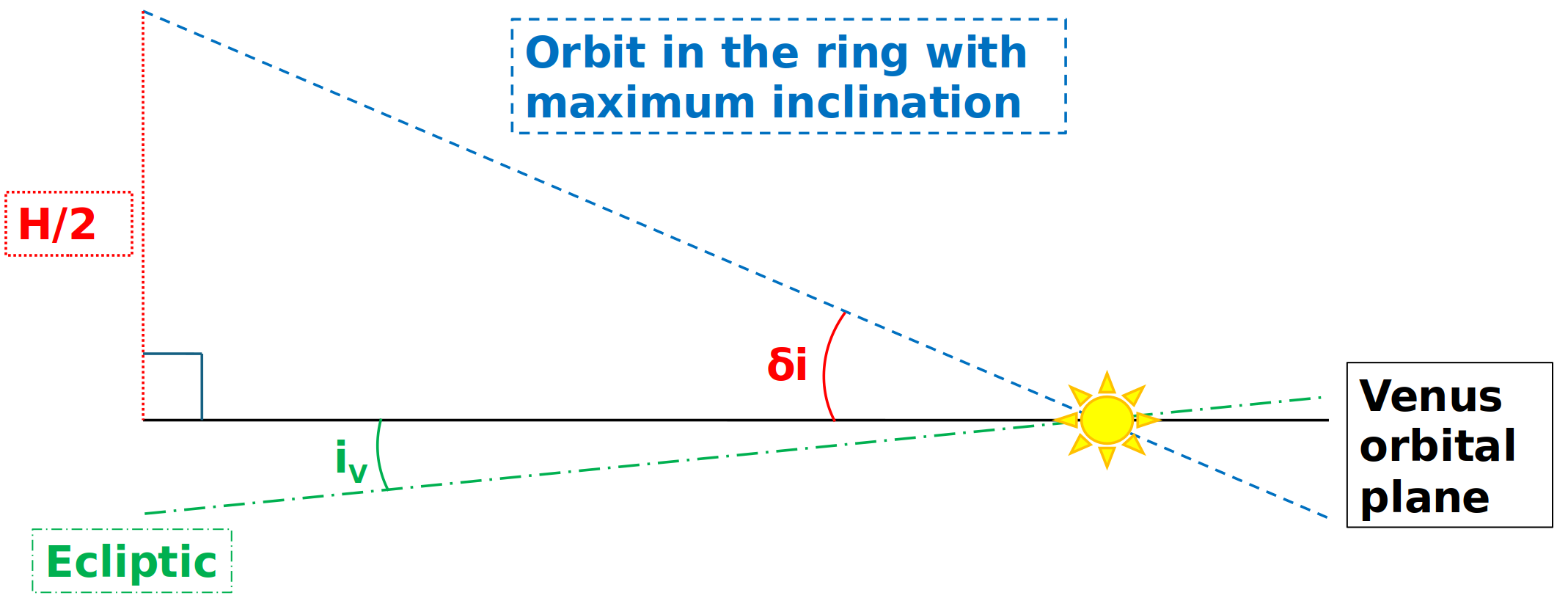}
    \caption{Schematics describing the computation of maximum inclination. The horizontal black line represents the orbital plane of Venus (at inclination $i_V$, in green), the dash-dot green line is the ecliptic, and the dashed blue line represents the orbit in the ring with the maximum inclination possible. The limit in height of the ring is represented through the dotted red line. A simple trigonometric equation allowed us to compute $\delta i$, the red angle. None of the angles represented here are to scale with the actual values.}
    \label{fig:incl}
\end{figure}

For the inclination, we chose a different approach. It is indeed possible to compute a theoretical $\delta i$ that should be added or removed to Venus's inclination ($i_V = 3.39471\degree$) to reach the height of the ring. In other words, the aphelion of an orbit defined by the same parameters as Venus's orbit but with an inclination of $i_V + \delta i$ would reach, at maximum, half of the height of the ring compared to Venus's orbital plane. This definition let us compute the $\delta i$ thanks to a simple trigonometric equation (see Fig.~\ref{fig:incl}). We followed this approach, but we added a 50\% margin to the value of $\delta i$ in this way: $i_V \pm 1.5 \delta i$.

Finally, the mean anomaly was chosen, as in the first model, between 0 and $360\degree$. The intervals chosen are reported in Table~\ref{tab:orb_elts}.

Each particle generated is tested to verify if the first period is inside the ring. In the case it does not, the particle is discarded and another one is generated. 99999 particles were created, a third with a radius of 10~mm, a third with a radius of 1~mm, and the last third with a radius of 0.1~mm. The results of the integration from IC2 are described in the next section. These results include all particles from IC2, without filtering on whether they belong to the ring after the integration. This filtering is only necessary for Sect.~\ref{sec:density_risk}. It should be noted here that no particles left the Solar System or encountered the Sun, but 4270 crashed in Venus.

\section{Distribution, velocity, and evolution of particles in the ring}\label{sec:res}

To analyse the dust ring, we define two quantities for each particle: $r$ is the distance between the centre of the Sun and the particle and $h$ is the normal distance between the particle and Venus's orbital plane. These two quantities are enough to plot the profile of a cross-section of the ring. We are also interested in the mean anomaly of the particles, used as a proxy to describe the azimuthal variation of the ring.

\begin{figure}
    \centering
    \includegraphics[scale = 0.6]{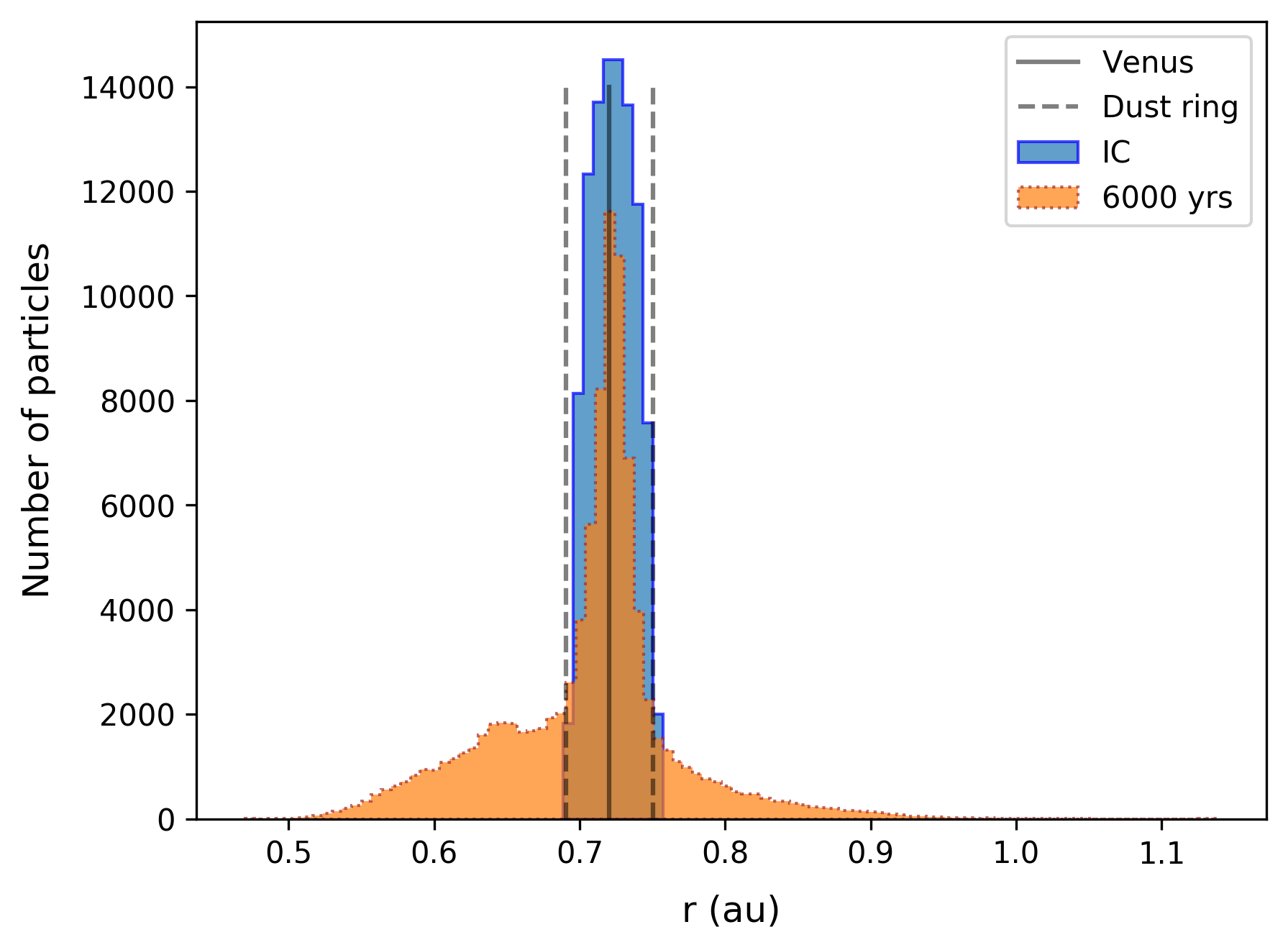}
    \caption{Histogram of the distribution of the particles on the $r$ values. The data in blue (`IC'; solid outline) are the initial conditions, meaning they represent the particles before the integration (IC2). The data in orange (dotted outline) represent the same particles but 6000~years later. The black line in the middle represents Venus at the end of the integration. The dashed lines around it underline the theoretical limits of the ring, as described in Sect.~\ref{sec:data}.}
    \label{fig:evol_r}
\end{figure}

\begin{figure}
    \centering
    \includegraphics[scale = 0.6]{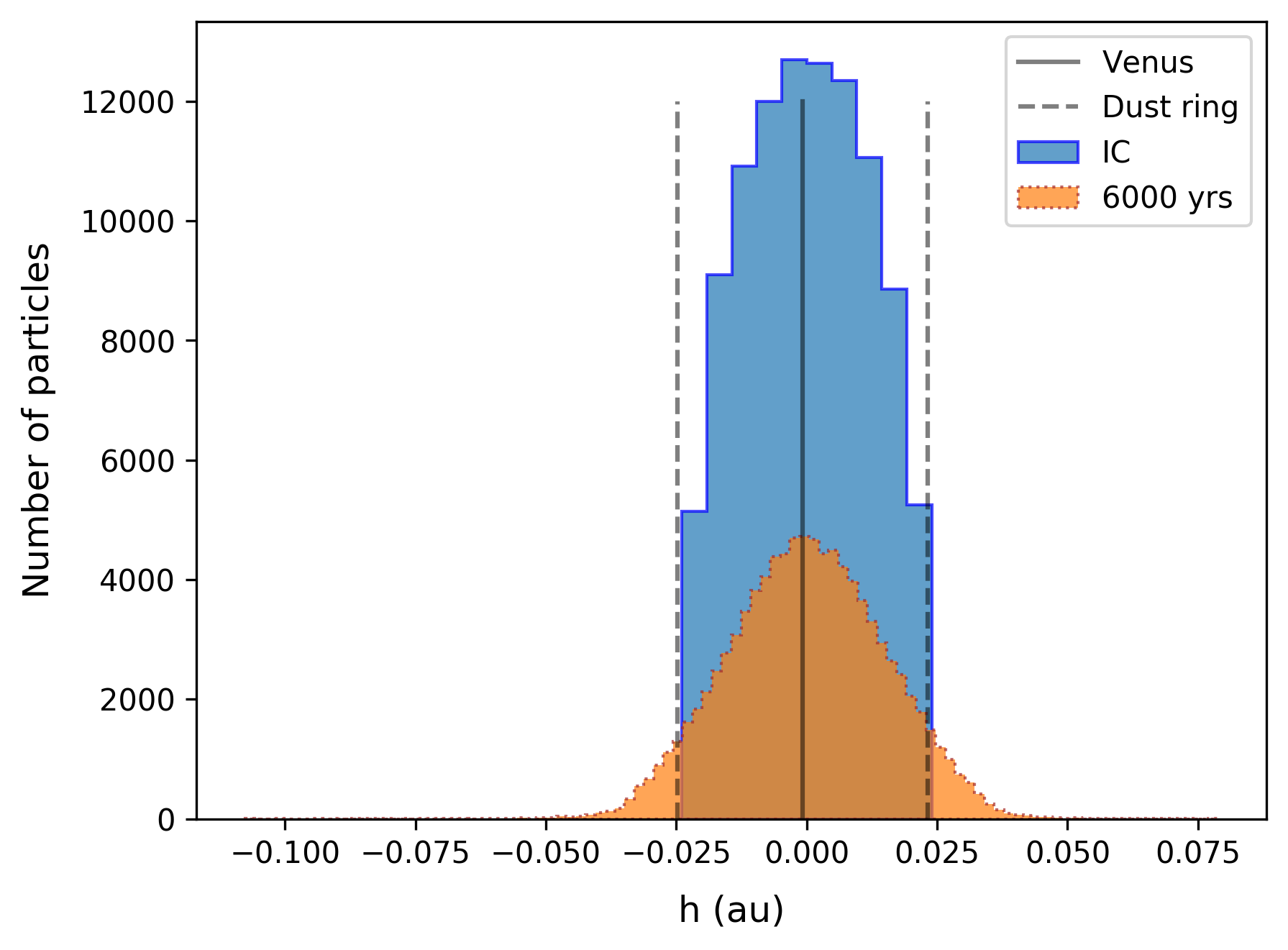}
    \caption{Histogram of the distribution of the particles on the $h$ values. This figure is similar to Fig.~\ref{fig:evol_r} and uses the same legend.}
    \label{fig:evol_h}
\end{figure}

\begin{figure}
    \centering
    \includegraphics[scale = 0.6]{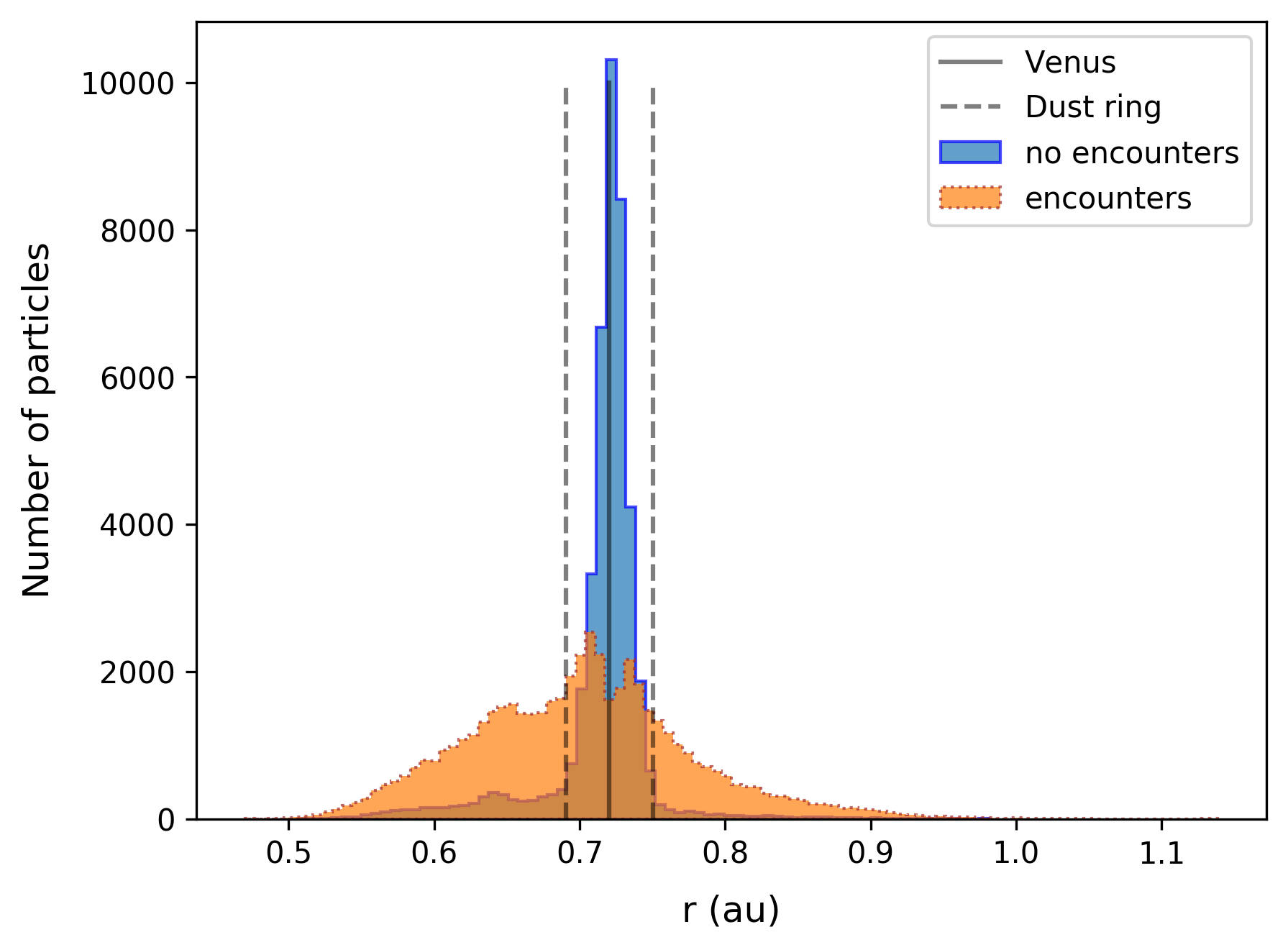}
    \caption{Histogram of the distribution of the particles on the $r$ values at the end of the integration. In blue (solid outline) is the distribution of particles that never encountered Venus, while in orange (dotted outline) is the distribution of particles that encountered Venus at some point during the integration. }
    \label{fig:evol_r_enc}
\end{figure}

Figures~\ref{fig:evol_r} and \ref{fig:evol_h} describe the distribution of the particles inside the ring, following $r$ and $h$ respectively, as well as the evolution of this distribution. The initial conditions of the particles fit nicely inside the ring as described from the observations (see Sect.~\ref{sec:data}). After 6000~years of integration, the ring has spread out, reaching $r = 0.47$~au at minimum and as far as 1.14~au, even though the amount of particles in these regions is much smaller. This seems to be due in majority to close encounters with Venus, as it can be seen in Fig.~\ref{fig:evol_r_enc}. Some particles (156) encountered the Earth, but these encounters did not have any impact on the r-distribution.

This spreading out of the dust ring is also observed in the distribution of particles along the $h$ values. Once more, the initial conditions respect the data observed, but the ring tends to spread out, although much less than in $r$.

In both figures, we notice the high concentration of particles at Venus's orbit, both in $r$ and in $h$. This concentration seems to come directly from our model, although it was not intended this way. However, the integration does not destroy this feature, so it may be a feature of the ring itself. We also confirm that the distribution centred on Venus's orbit holds very well over time, as particles on the outside of the ring form another Gaussian centred on Venus's orbit after the integration. Particles on the inside of the ring behave the same way. By `outside' (respectively `inside') we mean any particles whose $r$ is higher (lower) than Venus's $r$. This holds true as well for similar definitions using $h$.

\begin{figure}
    \centering
    \includegraphics[scale = 0.6]{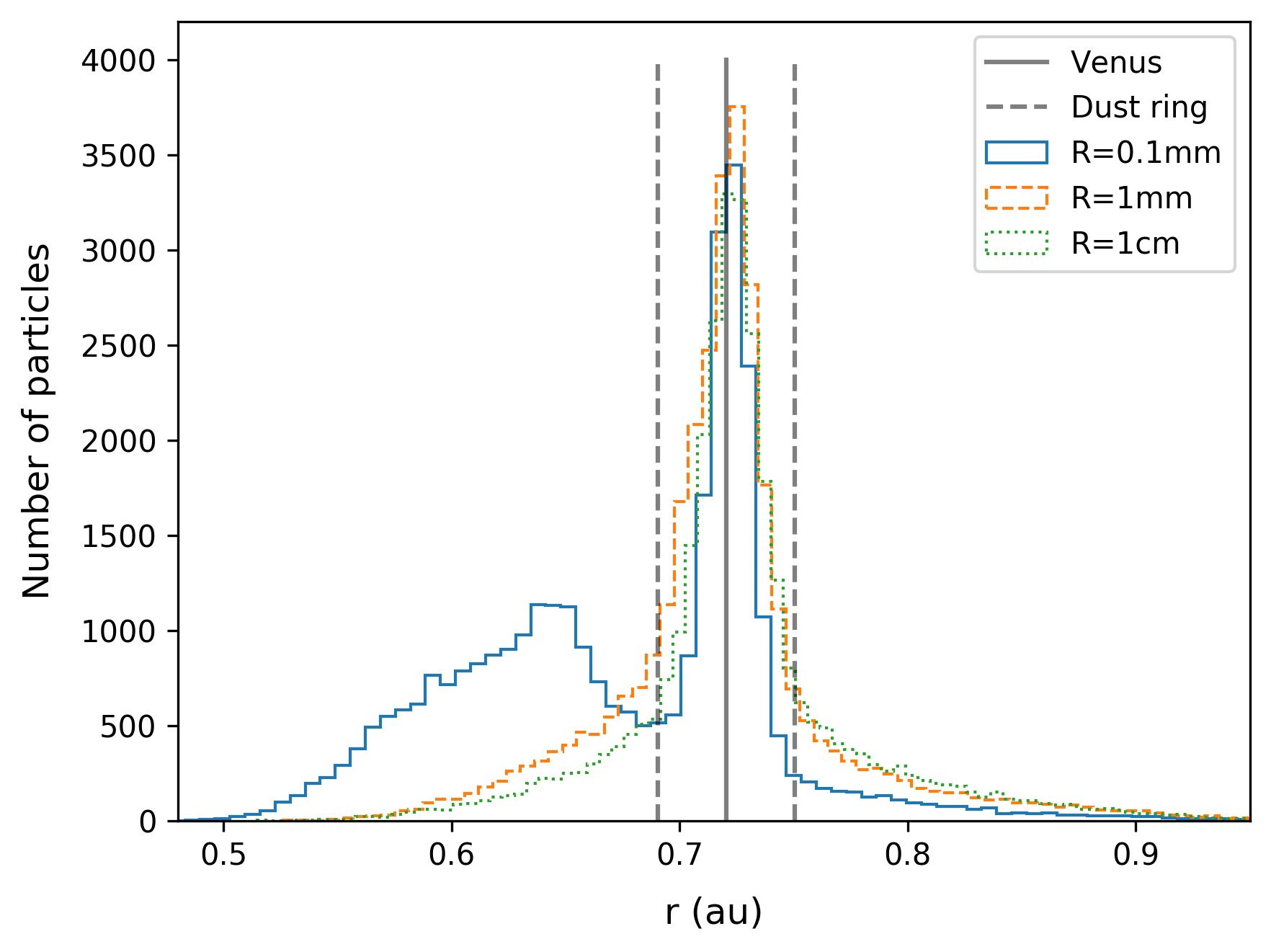}
    \caption{Histogram of the distribution of the particles on the $r$ values after 6000~years of integration, slightly zoomed-in. The black line represents Venus's orbit. The dashed grey lines show the limit of the dust ring. Each curve corresponds to a different particle size (dotted green line for 1~cm radius, dashed orange line for 1~mm, and solid blue line for 0.1~mm).}
    \label{fig:r_plr}
\end{figure}

 Interestingly, a very specific feature can be seen at around $r = 0.65$~au. It seems particles tend to concentrate at this point. One clue as to why comes from Fig.~\ref{fig:r_plr}, representing the distribution of particles in $r$ for different radii. Only the smallest particles follow this trend, while the other ones do not, which means the cause is NGFs. Indeed, the Poynting-Robertson drag in particular makes the smallest particles spiral towards the Sun. The peculiar peak shape we observe is only due to the initial peak distribution of the smallest particles, slowly moving towards a smaller distance to the Sun. We confirmed this by plotting the same figure for several integration time: a smaller peak starts to appear and moves slowly towards smaller $r$. As for the distribution in $h$, differences according to radii are negligible.

\begin{figure}
    \centering
    \includegraphics[scale = 0.6]{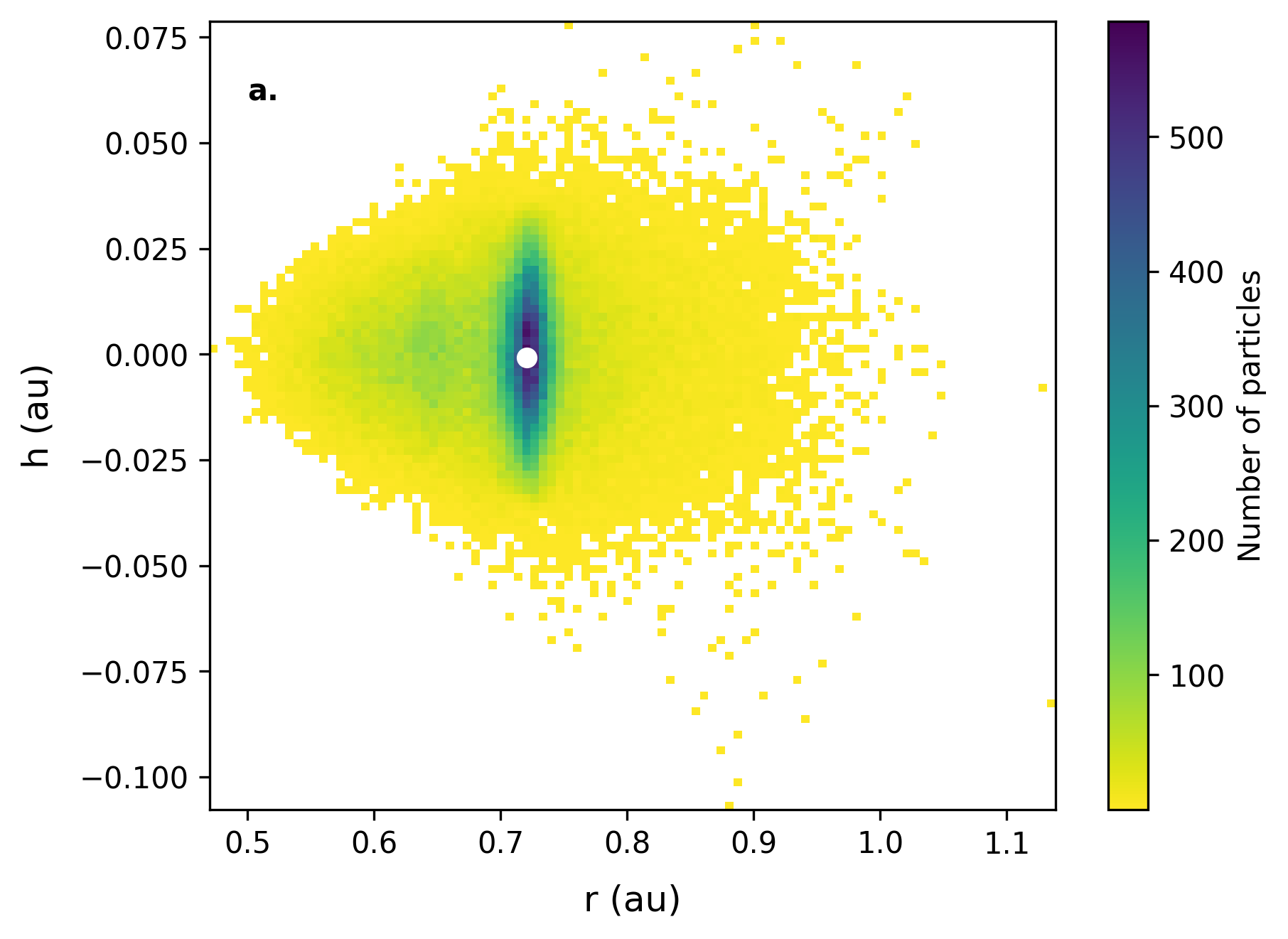}
    \includegraphics[scale = 0.6]{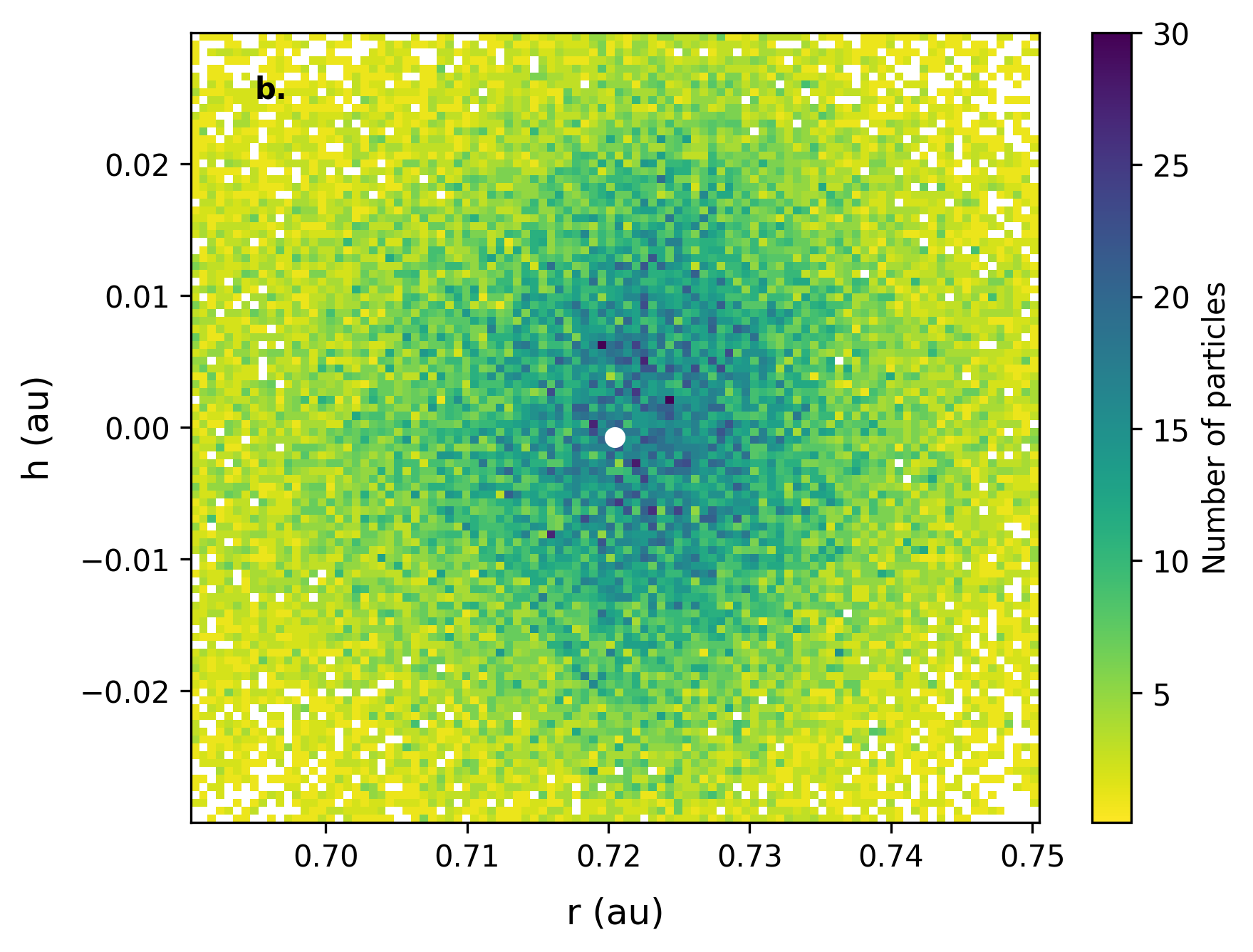}
    \caption{Two-dimensional histogram of particles after 6000~years of integration representing the cross-section of the ring. The white dot in the centre represents Venus's position at the end of the integration. Panel a: Representation of the whole section. Panel b: Similar 2D histogram, but zoomed-in on the centre of the ring, with $r$ and $h$ scaled similarly.}
    \label{fig:hist2d}
\end{figure}

The distribution of particles in a mean cross section can be represented by a 2D-histogram, in $r$ and $h$, drawn in Fig.~\ref{fig:hist2d}. The density of particles on Venus's orbit and close to it is very high before decreasing rapidly. Thus the risk for spacecraft seems at its greatest when they are following Venus's orbit but should decrease rapidly if they avoid the exact orbit of Venus. A more precise computation of the impact risk for spacecraft follows in Sect.~\ref{sec:density_risk}. The middle of the ring is more or less a disc centred on Venus, as can be seen in Fig.~\ref{fig:hist2d}.b. The feature noted in the $r$-histogram is also visible in Fig.~\ref{fig:hist2d}.a.

\begin{figure}
    \centering
    \includegraphics[scale = 0.6]{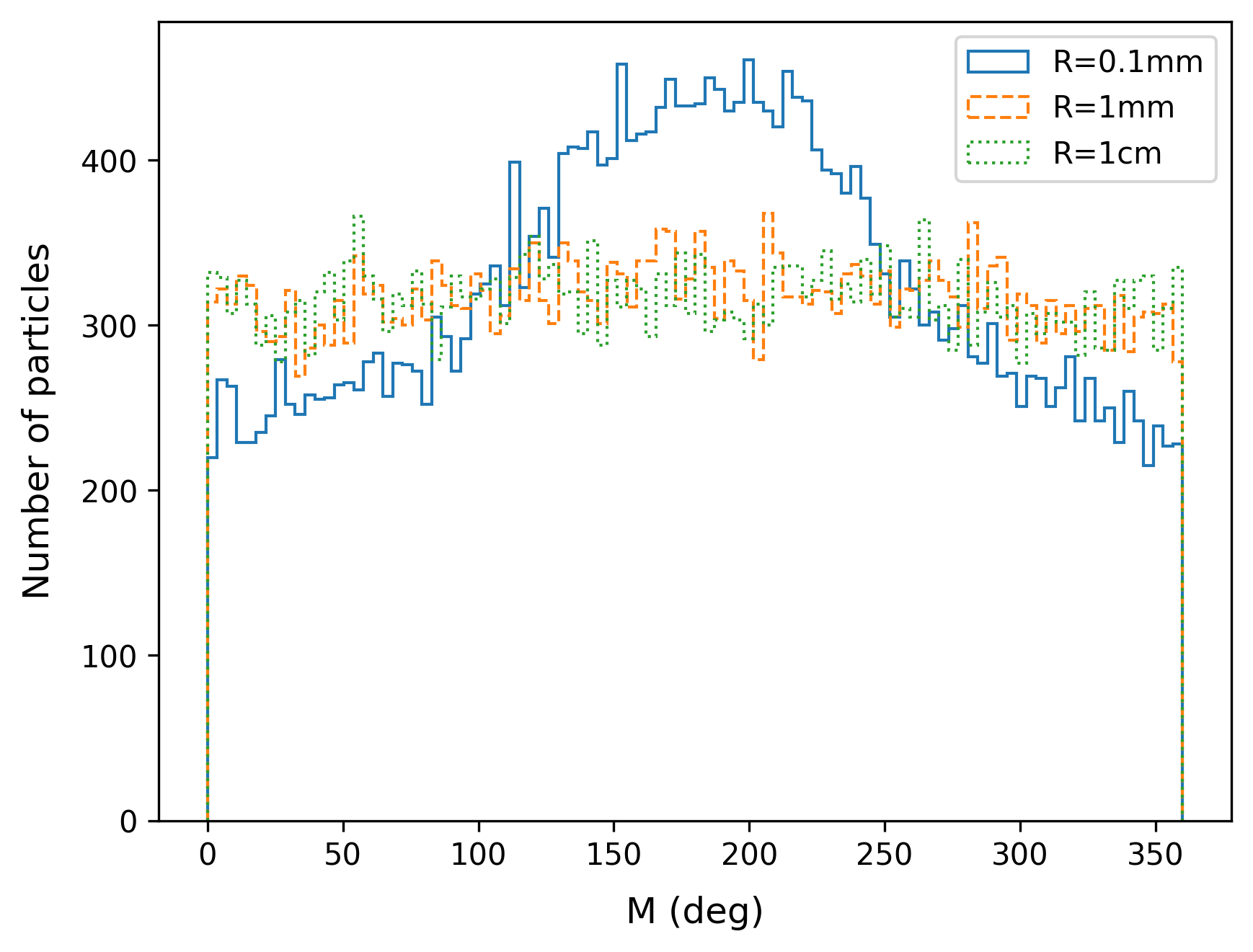}
    \caption{Histogram of the distribution of the particles on mean anomaly values (100 bins) after 6000~years of integration. Each curve corresponds to a different particle size (dotted green line for 1~cm radius, dashed orange line for 1~mm, and solid blue line for 0.1~mm).}
    \label{fig:hist_M_plr}
\end{figure}

The distribution of the particles azimuthaly is of course perfectly uniform initially since we chose the values of the mean anomaly between 0 and 360$\degree$ thanks to a uniform distribution. At the end of the integration, however, a bump in the histogram is visible. This bump is only due to the smallest particles, as it can be seen in Fig.~\ref{fig:hist_M_plr}. It does not seem to be linked to the position of Venus, as we observe it at different times without seeing a correlation with Venus's position. The resonant angle does not exhibit this feature, which appears only in the ring itself. It is probably linked with the NGFs, as they push the particles away where those forces have the strongest effect but let them concentrate where they have the weakest effect. The precise mechanism for this is unknown at this time, but investigating it further is beyond the scope of this paper. Because this feature is not too pronounced, we decided to neglect it in the rest of this study. 

\begin{figure}
    \centering
    \includegraphics[scale = 0.6]{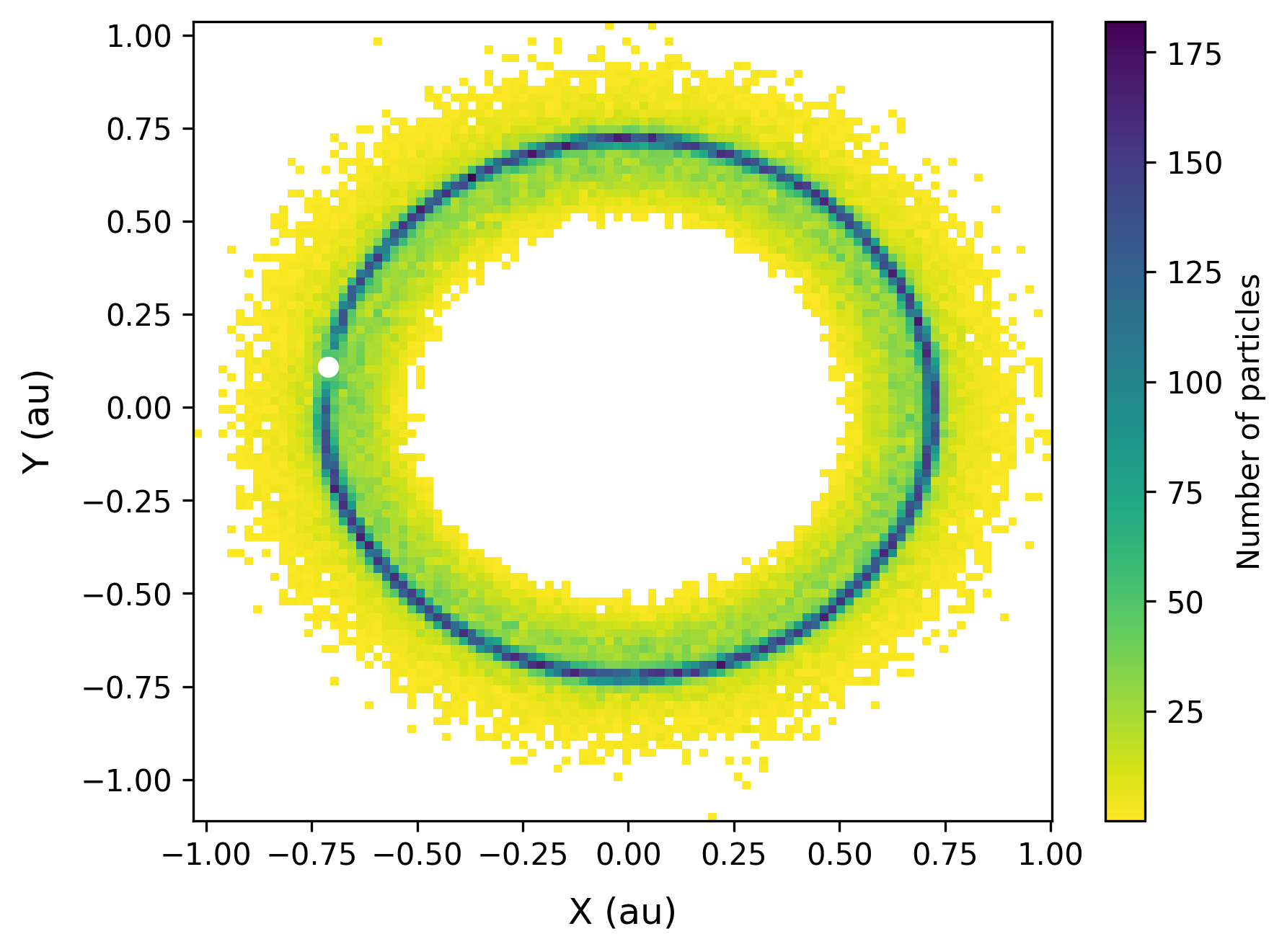}
    \caption{Histogram of the distribution of particles at the end of the integration according to X and Y values in the J2000 heliocentric reference frame defined in Spice. The white dot represents Venus's position.}
    \label{fig:hist_XY}
\end{figure}

Finally, Fig.~\ref{fig:hist_XY} shows the particles as seen from above at the end of the integration. The ring is clearly visible, with the highest concentration of particles on Venus's orbit. This concentration is only interrupted exactly at Venus since particles tend to be pushed away from the planet or simply crash into it. There is a slightly higher concentration of particles inside the dust ring, which is due to the feature visible in the $r$-histogram.

These results depend on the set of initial conditions we have described in Sect.~\ref{subsec:second_model}. However, we do not think they have a strong influence on the results, as a small variation in initial width or height would be negligible compared to the evolution observed.

\begin{figure}
    \centering
    \includegraphics[scale = 0.45]{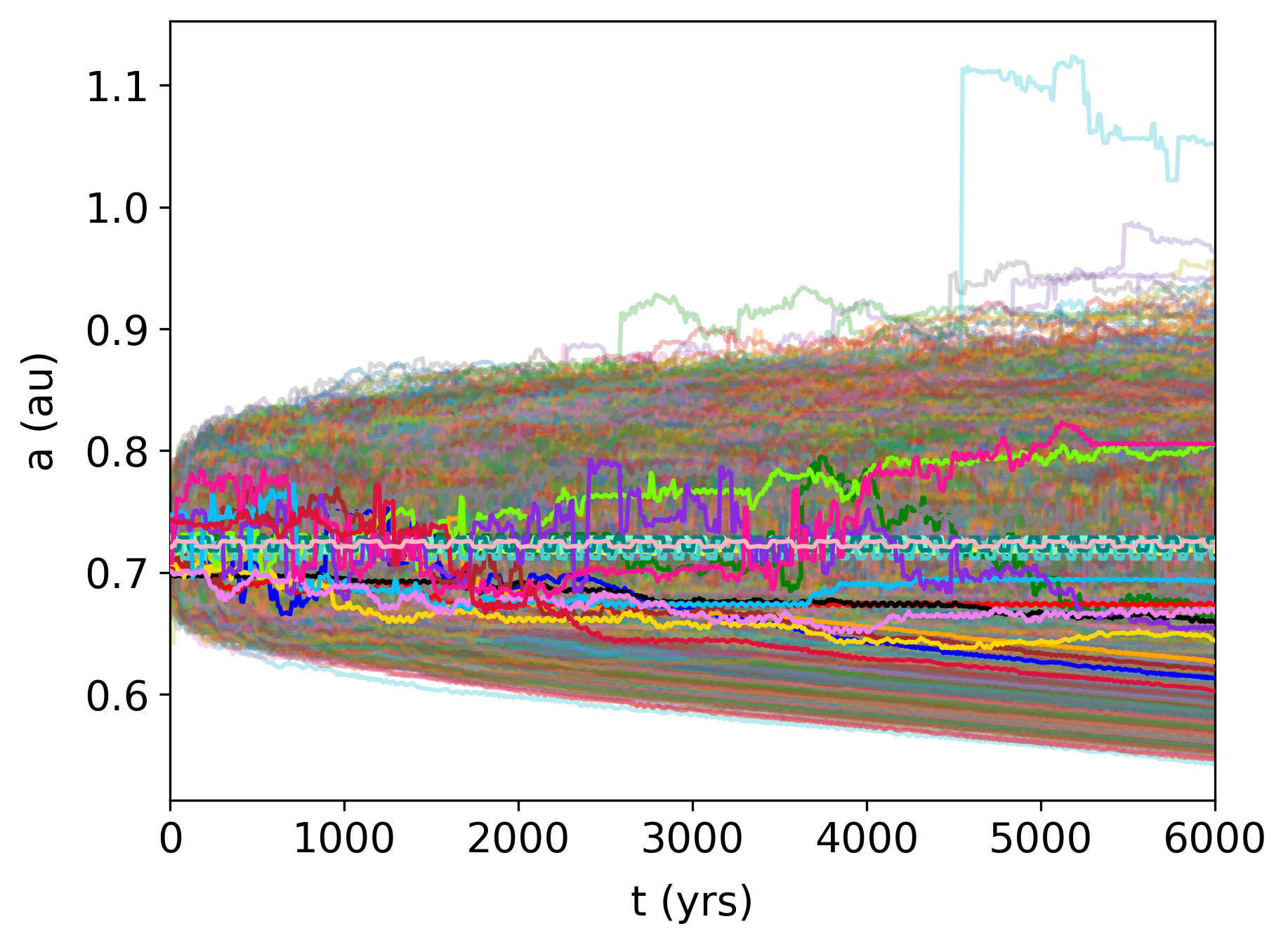}
    \includegraphics[scale = 0.45]{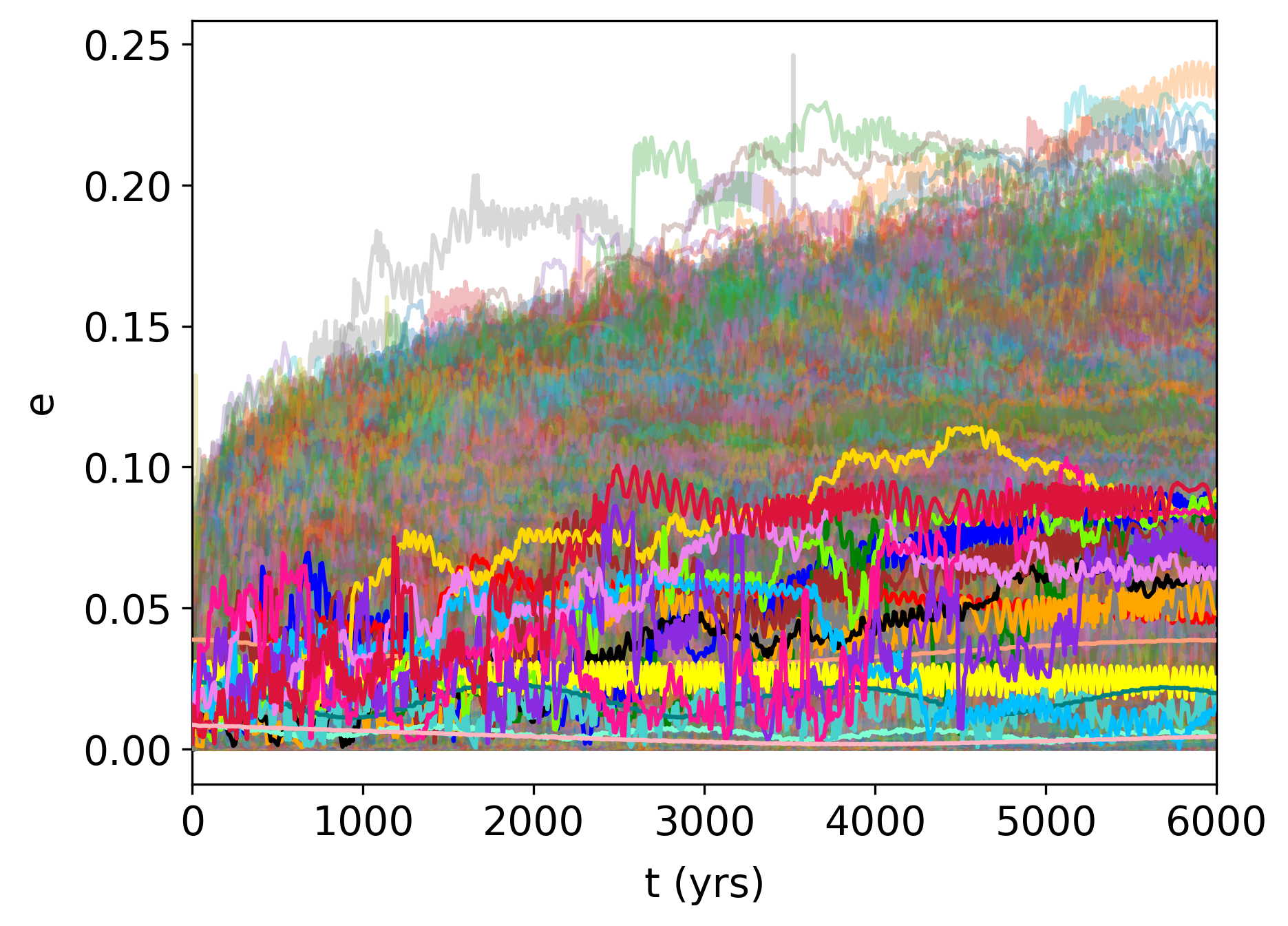}
    \includegraphics[scale = 0.45]{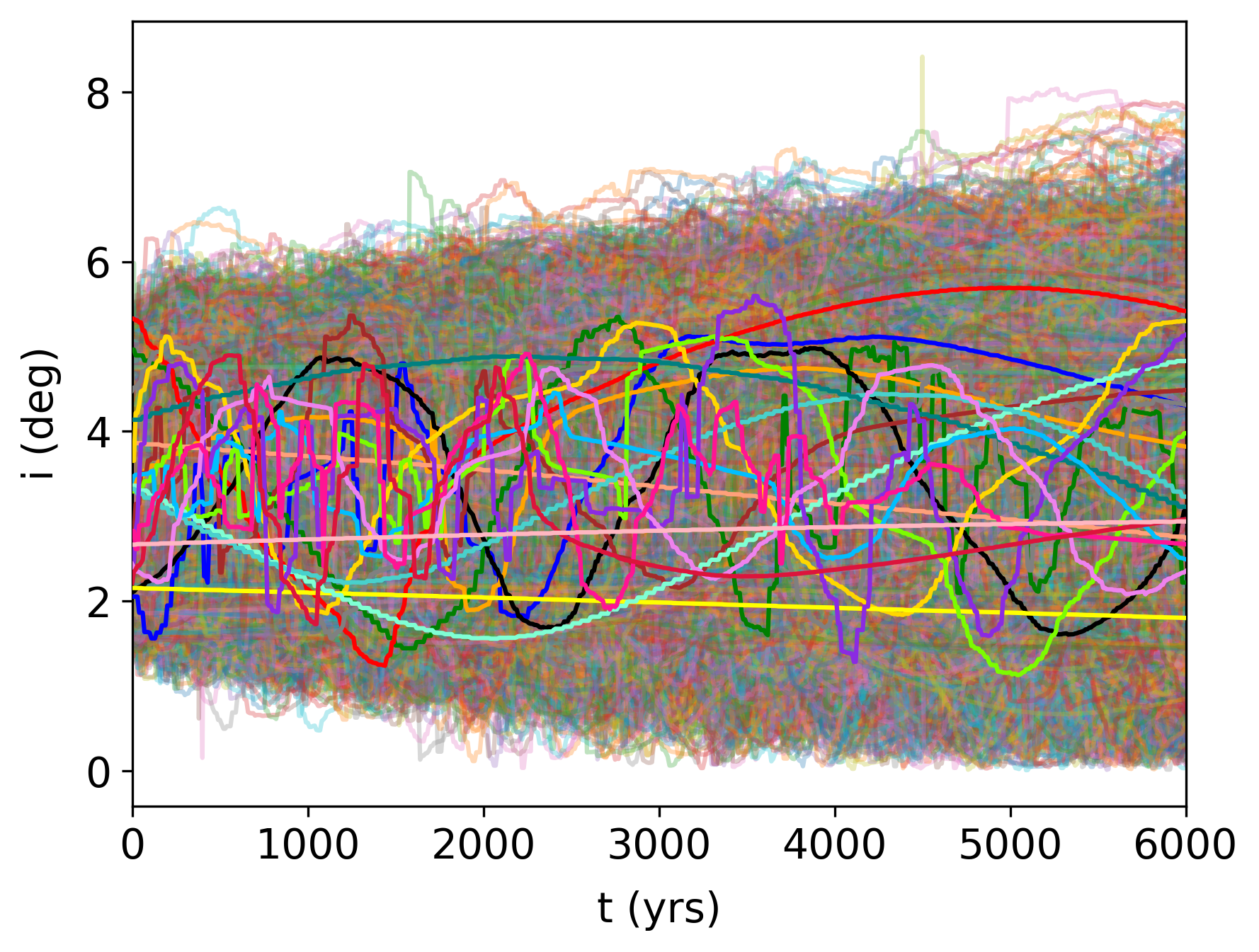}
    \includegraphics[scale = 0.45]{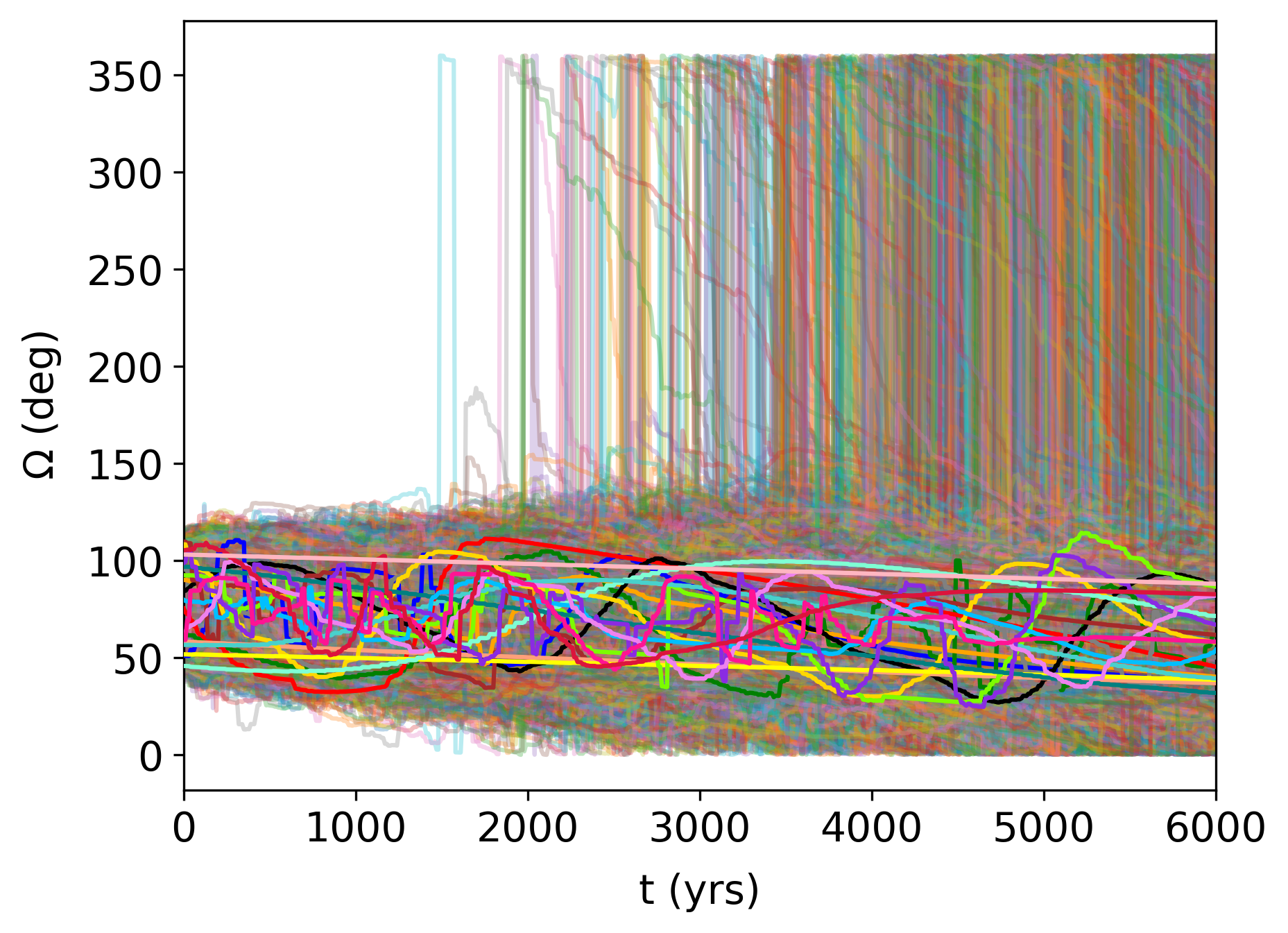}
    \caption{Evolution of four orbital elements (the semi-major axis, $a$; the eccentricity, $e$; the inclination, $i$; and the longitude of ascending node, $\Omega$) for all particles. Twenty randomly chosen particles are highlighted; the rest are represented to evaluate the envelope of the elements.}
    \label{fig:evol_elts}
\end{figure}

The evolution of the orbital elements can also be observed in Fig.~\ref{fig:evol_elts}. The envelopes of the semi-major axis, of the eccentricity, and of the inclination are still growing after 6000~years, showing that a complete picture of the ring would require longer timescales. However, we are only interested in selecting particles that best represent the ring in order to compute the impact risk. \citet{Pokorny_Kuchner_2019} show that the particles in the ring are stable over very long timescales. Thus, by selecting particles that are still in the ring after our 6000~years integration, we expect to obtain an accurate picture of the ring.

\section{Density of the ring and risk to a spacecraft: The example of BepiColombo}\label{sec:density_risk}

To properly evaluate the impact risk of the ring to a spacecraft, we evaluated the number of impacts of the meteoroids on the spacecraft, called the impact flux. To do this, we computed the number of particles inside the ring, as well as their mass and size distribution. We also evaluated the speed of impacts and the direction of the impacts, in order to investigate if the meteoroids preferentially hit sensitive parts of a spacecraft, such as solar panels.

This computation can only be done for a specific spacecraft since its trajectory has a major effect on the impact flux. We chose as an example BepiColombo, a spacecraft that crossed the dust ring region.

Here we describe the method to evaluate the speed and direction of impacting particles as well as the impact flux. First, we propagated BepiColombo in our model of the ring and recorded any impact with it. This is used to record the speed and direction of encounters. Then, having computed the number of particles we expect in the real ring, we scaled up the number of impacts to be able to deduce the impact flux. Afterward, we discuss the speed and direction of impact.

\subsection{Recording of impacts in our model}\label{sec:enc}

BepiColombo is composed of two modules: the Mercury Planetary Orbiter (European space agency - ESA) and the Mercury Magnetospheric Orbiter (Japanese space agency - JAXA). During the trajectory from the Earth to Mercury, these two modules are encased in another one: the Mercury Transfer Module (ESA). This is the configuration we are studying in this paper since BepiColombo crossed the Venus dust ring during this transfer phase.

Indeed, to reach Mercury, BepiColombo had to perform nine flybys, including two of Venus, on October 15th, 2020 (closest approach at 10720~km) and on August 10th, 2021 (closest approach at 552~km). These flybys mean that BepiColombo crossed the dust ring, although its effect on the satellite was not computed before the launch, as the ring was not incorporated in the impact models. 

\begin{figure}
    \centering
    \includegraphics[scale = 0.05]{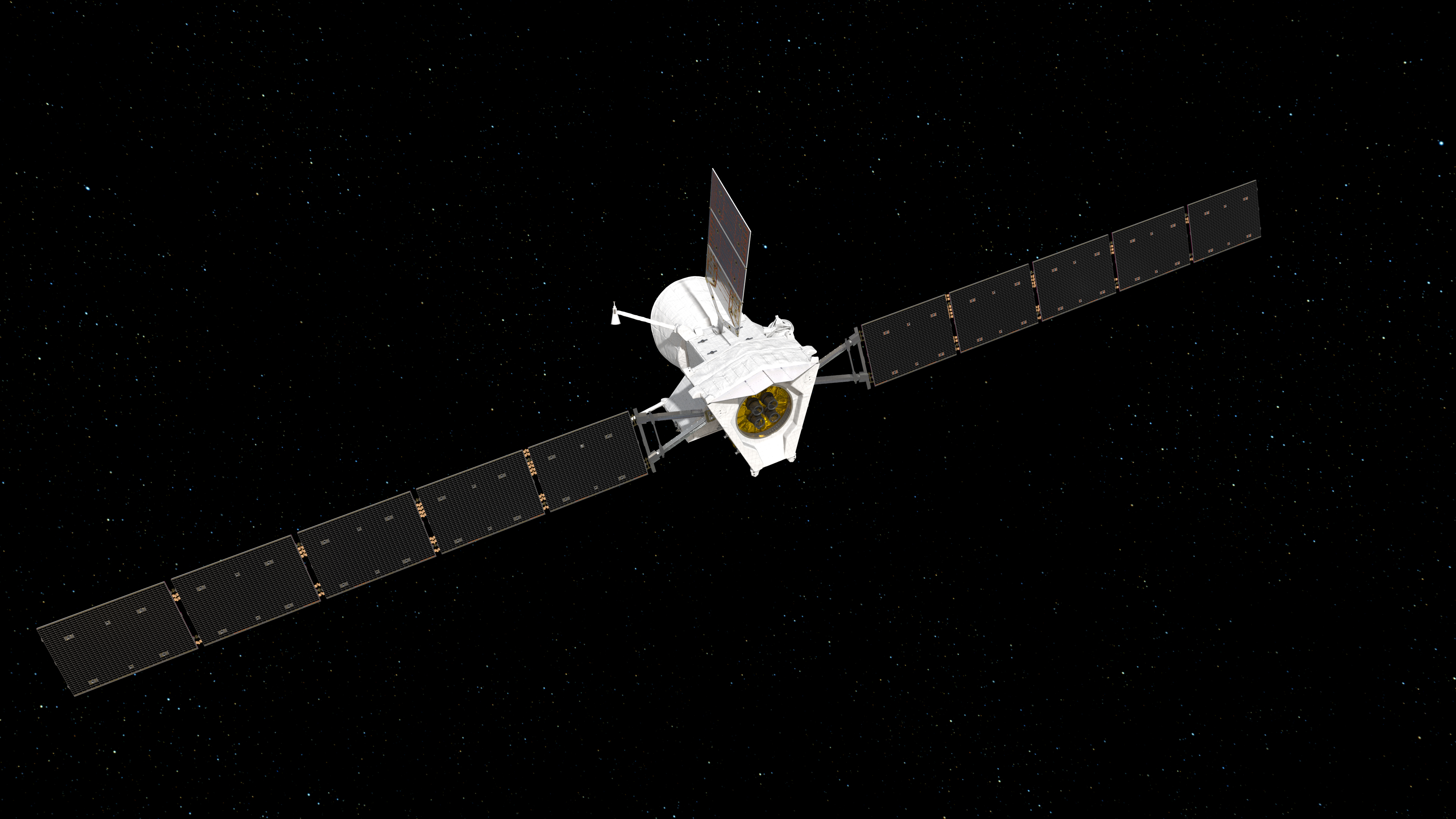}
    \caption{Artist impression of BepiColombo during the transfer phase towards Mercury (ESA).}
    \label{fig:bepi}
\end{figure}

There are no dust detectors on the Mercury Transfer Module and the one on the Mercury Magnetospheric Orbiter is in a casing during the transfer phase, which means we cannot use measurements from BepiColombo to constrain the dust ring. The module measures 30~m in width when its solar panels are deployed. They have a surface of 42~$\text{m}^2$ (see Fig.~\ref{fig:bepi}).

To record the impacts in our model, we first selected the orbits of the particles that still fit inside the ring after the integration. These orbits are the basis for the database that we use to evaluate the impact risks.

To check whether a particle is still inside the dust ring, we followed the same reasoning as when modelling the dust ring: A particle is considered as belonging to the dust ring if it stays inside it for one period. The equation describing this is Eq.~\ref{eq}. Only 27.28\% of particles satisfy this condition. The orbits corresponding to these particles are thus a good description of the dust ring: the particles stayed inside for 6000 years.

The dust ring contains 9494~particles of radius 10~mm, 9270~particles of radius 1~mm, and 7350~particles of radius 0.1~mm. This discrepancy between the smallest particles and the other sizes is simply due to the specific orbital evolution of the smallest particles, because of NGFs, as described in Sect.~\ref{sec:res}.

We made the hypothesis that the actual number of particles in the ring is greater than this (this is confirmed in the next section). To compute the impact risk, we multiplied the number of particles in the following way. For each orbit selected, we sampled the mean anomaly so that 1000 values could be fitted on the orbit, thus multiplying the amount of particles modelled by 1000. The only side effect is that we destroyed the feature in mean anomaly visible in Fig.~\ref{fig:hist_M_plr} that we had decided to neglect. This forms the database of particles we used to compute the impact risk.

We propagated BepiColombo in this database, following its transfer trajectory between the Earth and Mercury, and we marked the particles that are closer to BepiColombo than a distance $d$ for each time step. We chose $d$ = 0.005~au after several trials, as it is the lowest possible value that gives statistically meaningful results. For each particle encountered, we recorded the speed and direction of the particle and of BepiColombo. 

\subsection{Density of the ring and impact flux on BepiColombo}\label{sec:density}

With the method explained in the previous subsection, we obtained a number of impacts on BepiColombo. However, to evaluate the actual number of impacts, we needed to compute the number of particles in the ring, necessarily higher than in our model. Subsequently, we scaled the number of impacts up to the actual value.

The observations revealed that the ring density is 10\% superior to the number density in the surrounding interplanetary dust cloud. We used the tool IMEM2 \citep{Soja_al_2019} to compute the density on Venus's orbit. This density does not take the ring into account, because IMEM2 does not account for it, so we had to add the 10\%.

IMEM2 is based on the integration of particles ejected from three different types of parent bodies (Jupiter-family comets, Halley-type comets, and asteroids) for 1 million years. The resulting model was then fitted to data to adjust each parameter in order to obtain a good description of the dust environment from the Sun to the orbit of Jupiter. There are 12 sizes of particles, from 1~$\mu$m to 1~cm and three density bins, according to the origin of the particles.

\begin{figure}
    \centering
    \includegraphics[scale = 0.6]{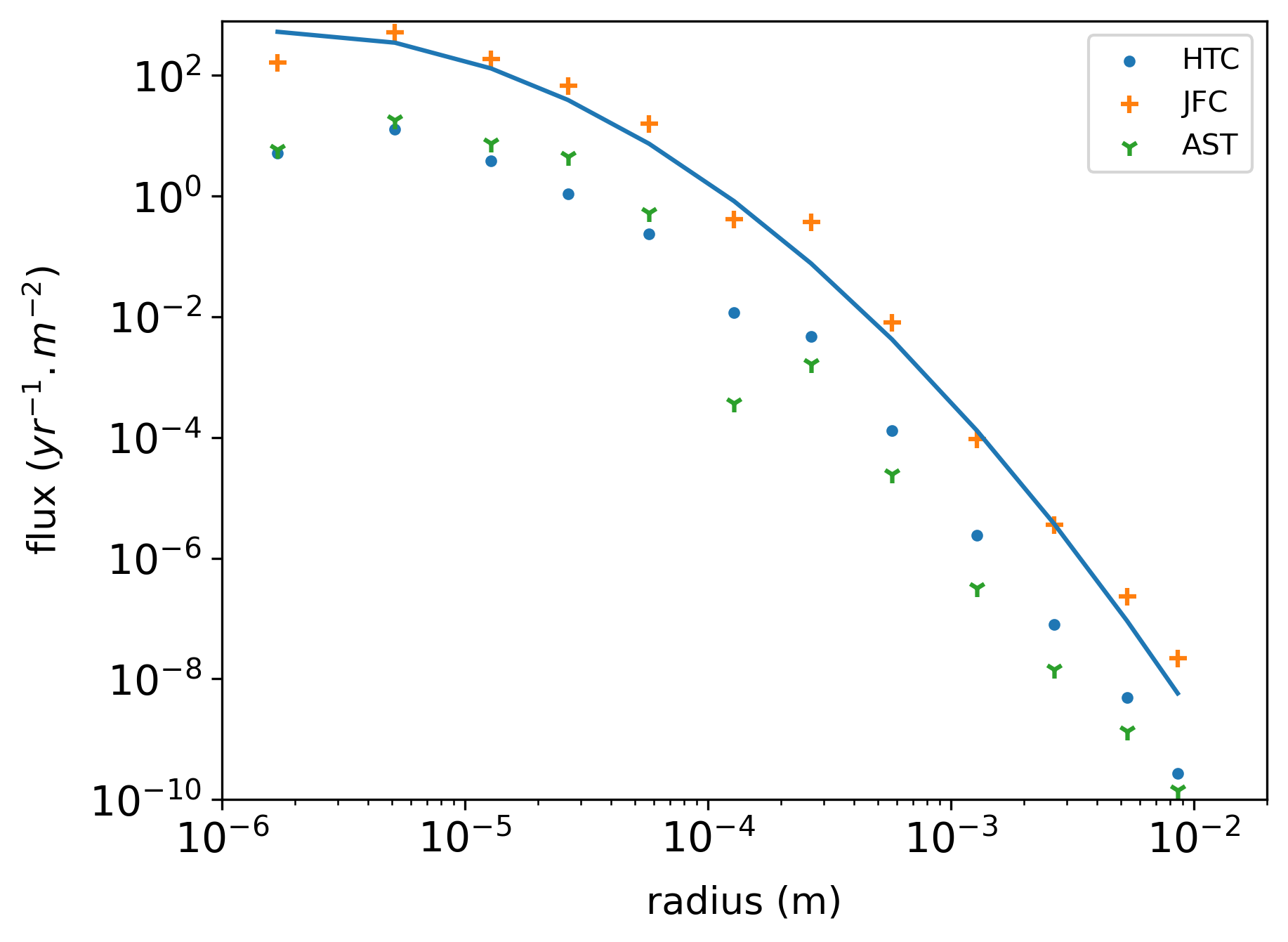}
    \caption{Impact flux from IMEM2 at Venus's orbit without the dust ring from three different sources (points in color) and a plot of the impact flux on BepiColombo from the dust ring with the extrapolation thanks to IMEM2 (solid blue curve), both as a function of the radius of the particles. The legend shows the origin of the particles according to IMEM2 (`HST': Halley-type comet; `AST': asteroids; `JFC': Jupiter-family comet). Fig.~\ref{fig:flux_final_rayon} shows more details of the impact flux on BepiColombo.}
    \label{fig:comp_bepi}
\end{figure}

Figure~\ref{fig:comp_bepi} shows the flux as computed by IMEM2 at Venus's orbit.  To compute the number of particles in the ring from the flux given by IMEM2, we wrote
\begin{equation}
    n (R) = 1.10 \times S_{ring} \times F(R)  \times T,
\end{equation}
with $n$ the number of particles in the ring, $F$ the flux given by IMEM2, which only depends on the radius $R$ of particles, and T the period of Venus. The number 1.10 takes into account the full density of the ring: 10\% more than the sporadic background, which makes up 100\% of IMEM2 result. $S_{ring}$ is the surface of the cross-section of the ring, approximated by an ellipse with semi-major axis $w/2$ and semi-minor axis $H/2$. The flux F is the sum of the three fluxes as a function of origins, given by IMEM2. 

We thus obtained the number of particles inside the ring for each mass bin. This number reaches very high values due to the very large volume of the ring: from $10^{9}$ for the largest radii ($1.62.10^{-14}$ particles per kilometre cube) to $10^{22}$ for the smallest ones ($0.162$ particles per kilometre cube). This validates our hypothesis.

Using this data and the recording of impacts from our model, we computed the impact flux on BepiColombo from the actual dust ring (with the real number of particles). To do this, we counted how many particles of each radius bin were impacted using our database. We then divided this number by the area of a sphere of radius $d$ and by the time BepiColombo spends close to the ring (about 1.77~years). Finally, we scaled these three data points to the actual number of particles in the ring. 

\begin{figure}
    \centering
    \includegraphics[scale = 0.6]{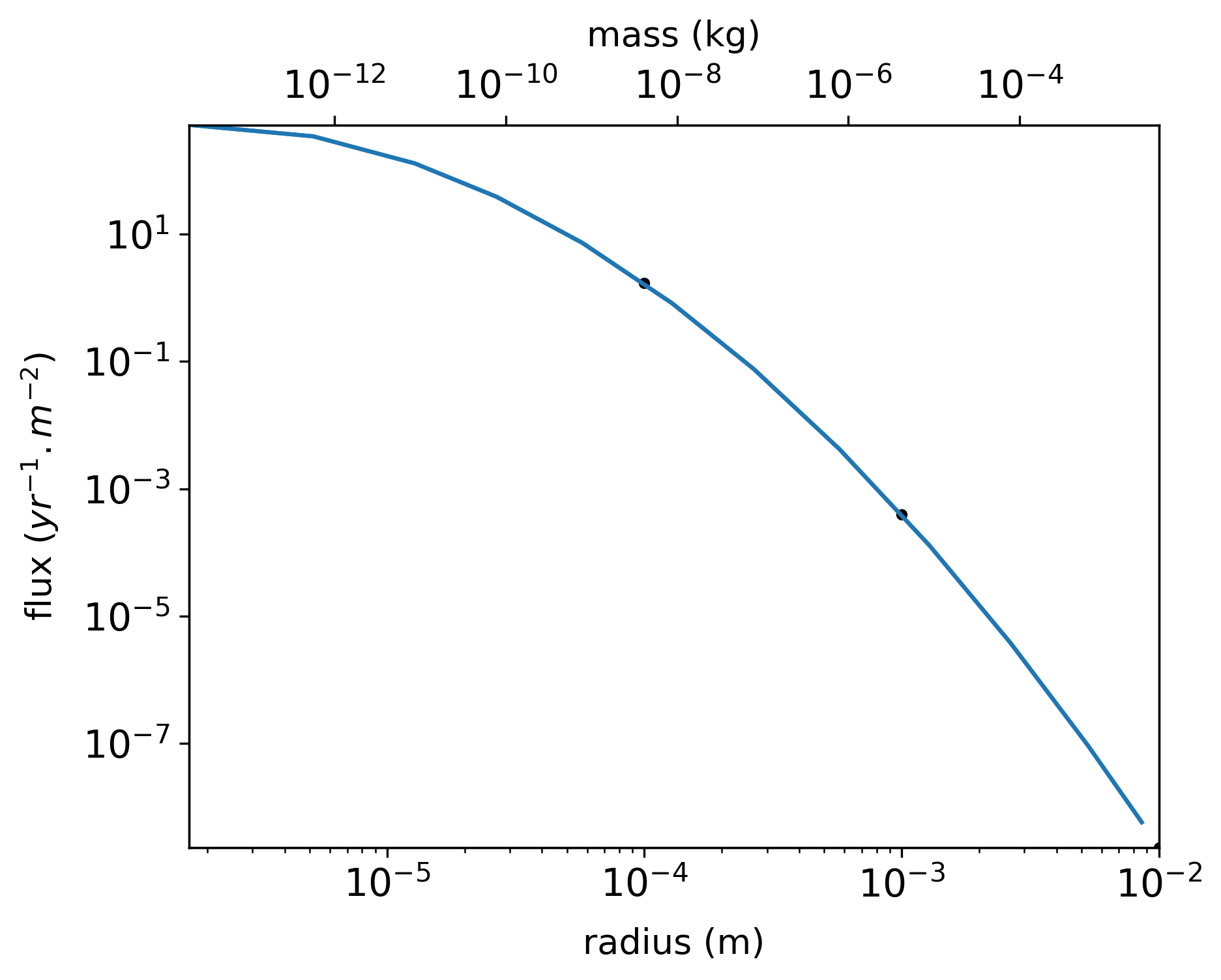}
    \caption{Impact flux on BepiColombo due to the dust ring as a function of radius and mass (see text). The blue line represents the function, $f$, fitted to the three data points (black dots) obtained from our model.}
    \label{fig:flux_final_rayon}
\end{figure}

To evaluate the impact flux for smaller radii than the ones we modelled, we used IMEM2. First, we fitted the flux points from IMEM2 with a quadratic function that takes as its argument the logarithm 10 of the radius and returns the logarithm 10 of the flux. We call this function $f$: $f = a.x^2 + b.x + c$, with $x = log_{10}(R)$, $y = log_{10}(F)$, $a = -0.802$, $b = -9.25$, and $c = -25.0$ (standard deviation of each parameter: 0.527,  1.12, and  2.02). Then we modified this function by fitting it on the three data points we already computed, adjusting only the value of $c$. We obtained $c = -23.9$. This is possible because we computed the three data points using the distribution from IMEM2. In this way, we broaden the results to the radii covered in IMEM2. Figure~\ref{fig:flux_final_rayon} shows the new function and the three data points, giving a general idea of the fluxes we expect in the dust ring along BepiColombo trajectory.

These fluxes can be compared to the reference for the interplanetary dust fluxes we expect in that region of space (see Fig.~\ref{fig:comp_bepi}). The fluxes we computed for BepiColombo are about the same order of magnitude than the IMEM2 reference, which means that these fluxes are comparable to the effect of the general meteoroid environment, which is not a major threat.

These fluxes show clearly that the biggest particles (10~mm) has almost no chance of being encountered (a satellite of cross-section of 100~$\text{m}^2$ would need on average about $10^6$~yrs to encounter such a particle), which means that the risk of being destroyed by a big particle because of the Venus dust ring is negligible. However, there are much more chances of encountering smaller particles. This should be kept in mind for the rest of this discussion. 

We recomputed the impact fluxes on BepiColombo, but as a function of mass (see Fig.~\ref{fig:flux_final_rayon}), as it is most often how impact fluxes are studied. To do this, we simply computed the mass from the radius using the bulk density chosen in our model (1000~kg/$\text{m}^3$). 

\begin{figure}
    \centering
    \includegraphics[scale = 0.6]{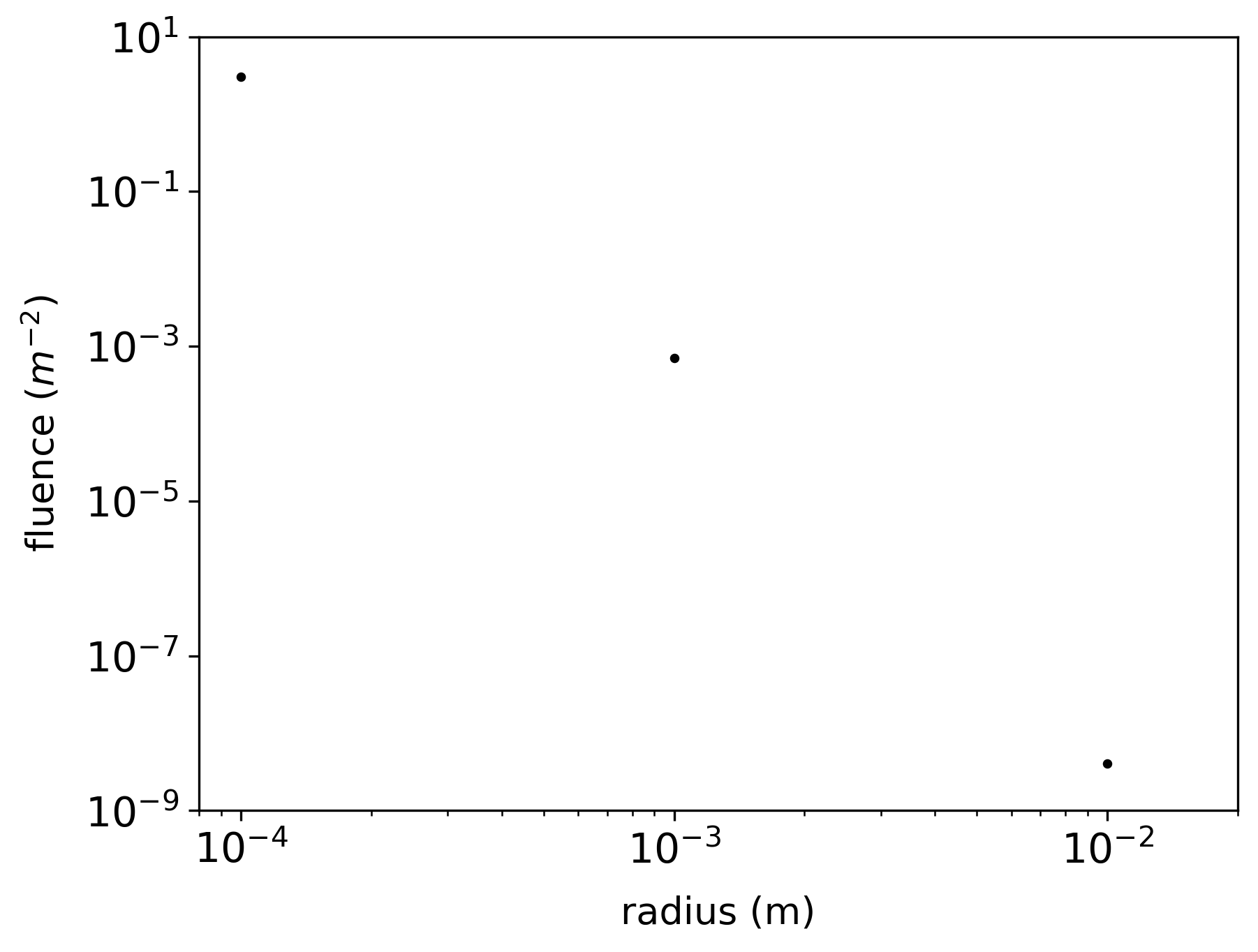}
    \caption{Fluence from the dust ring on BepiColombo on the duration of the mission}
    \label{fig:fluence}
\end{figure}

Finally, we compute the fluence from the dust ring on BepiColombo. We take into account the whole duration of the transfer between the Earth and Mercury's orbit (seven years). Figure~\ref{fig:fluence} shows the risk to the spacecraft from the dust ring. A few particles per square meter will impact the spacecraft during the transfer, but all of the particles are very small. This shows how the short time spent in the dust ring helps minimise the risk.

In conclusion, we have shown that the dust flux and fluence from the dust ring are not high enough to be considered dangerous to BepiColombo. Using both the number of impacts from our model and extrapolation from IMEM2, we have also shown that impacts are mostly with the smallest particles, while the biggest ones are almost never encountered.

\subsection{Speed and direction of impact}\label{sec:speed_dir}

In this subsection we discuss the results from the encounters recorded when propagating BepiColombo inside our model of the dust ring. To have a meaningful discussion on this topic, we chose to plot the results as fluxes, not number of impacts. This means we use the same reasoning as the previous section to compute the fluxes we present here, scaling them according to amount of particles in the ring and using IMEM2 as reference.

\begin{figure}
    \centering
    \includegraphics[scale = 0.55]{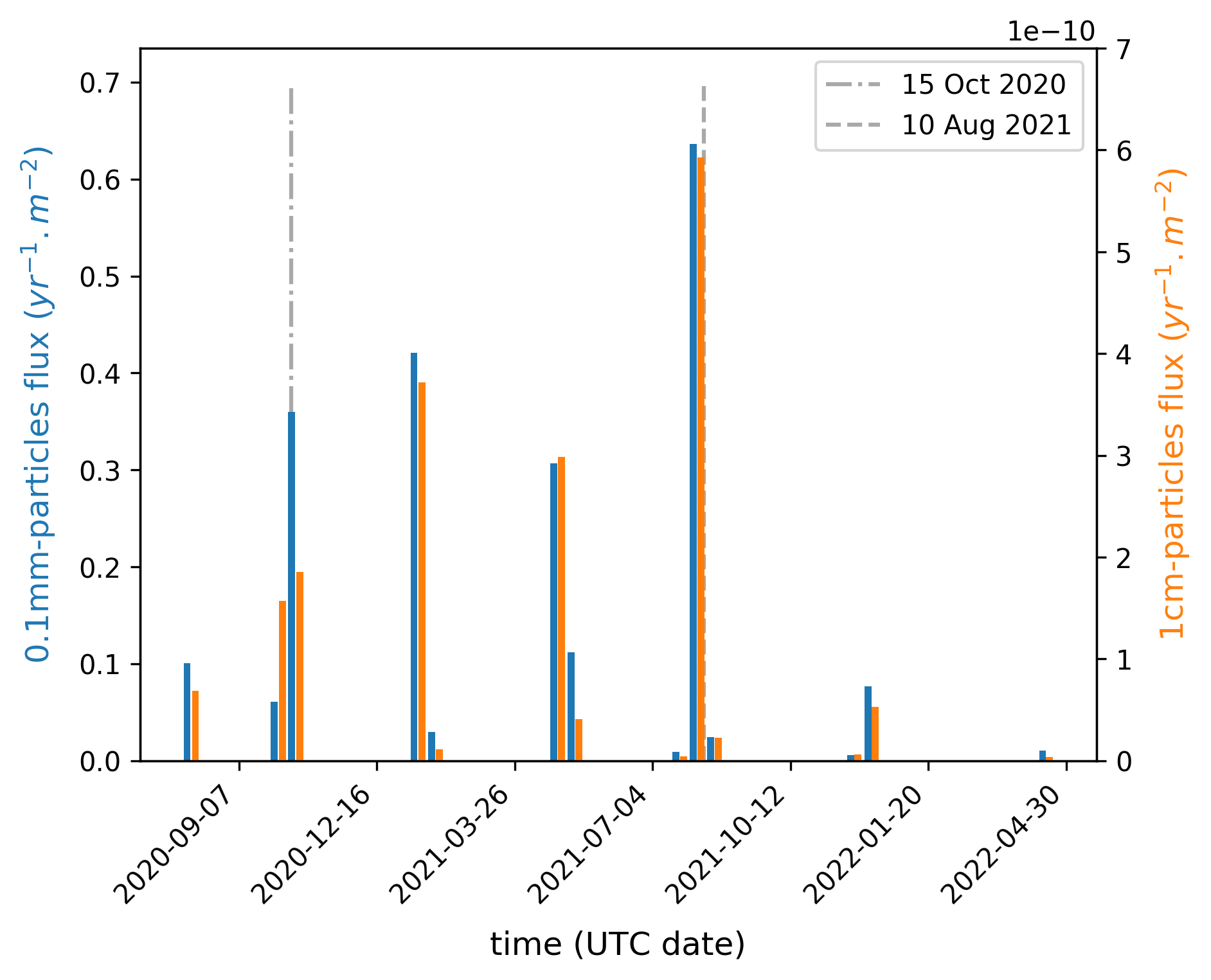}
    \caption{Evolution of impact flux from the dust ring on BepiColombo as a function of the date of impact. Each y-axis represents the impact flux of a specific particle size (0.1mm in blue, on the left and 1cm in orange, on the right). The two dates marked with a grey line are the dates of the two flybys of Venus by BepiColombo, the second one having the closest approach.}
    \label{fig:impact_date}
\end{figure}

Figure~\ref{fig:impact_date} shows the flux over time, for two sizes of particles. The third size shows very similar results. There are four major peaks, two correspond perfectly to the gravity assist maneuvers by BepiColombo (15 October 2020 and 10 August 2021). This second date corresponds to the closest approach of Venus by BepiColombo, which translates to the highest peak in the distribution. 

\begin{figure}
    \centering
    \includegraphics[scale = 0.55]{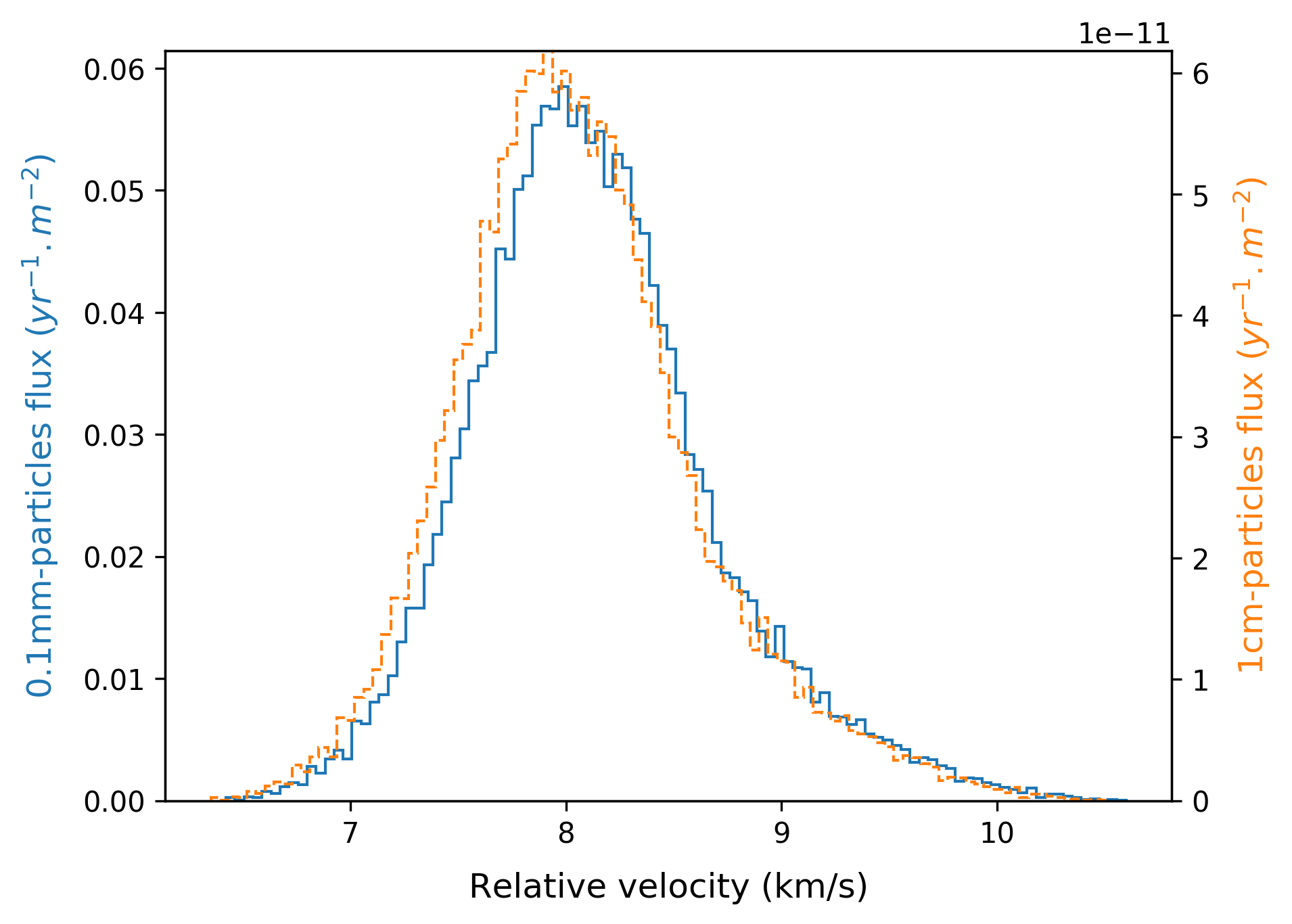}
    \caption{For two of the radii modelled, the distribution of the impact flux from the dust ring on BepiColombo as a function of the norm of the relative velocity of the particles relative to BepiColombo. Each particle size has a specific y-axis (blue, left for the 0.1mm particles and orange, right for the 1cm particles) }
    \label{fig:impact_speed}
\end{figure}

Figure~\ref{fig:impact_speed} shows the distribution of the impact flux as a function of relative velocity. The velocities reach as high as 11~km/s, but the peak in the distribution falls close to 8~km/s. Keeping in mind that impacts mostly happen with the smallest particles and rarely with the biggest ones, as previous computations have shown, the velocity values are low and not dangerous to the spacecraft, as a quick computation of the impact energy would show. A good point of reference for the most probable meteoroid velocities is around 15-20~km/s \citep{Soja_al_2019}. 

Since both the impact flux and the speed of impact are not very high, the risk can only come from another source: if there is a high probability of particles hitting the solar panels. These are indeed crucial to the functioning of the spacecraft. Our computation is not precise enough to detect if a particle could hit a solar panel directly, but we can analyse the direction of the encounters recorded to answer this question.

\begin{figure}
   \centering
    \includegraphics[scale = 0.5]{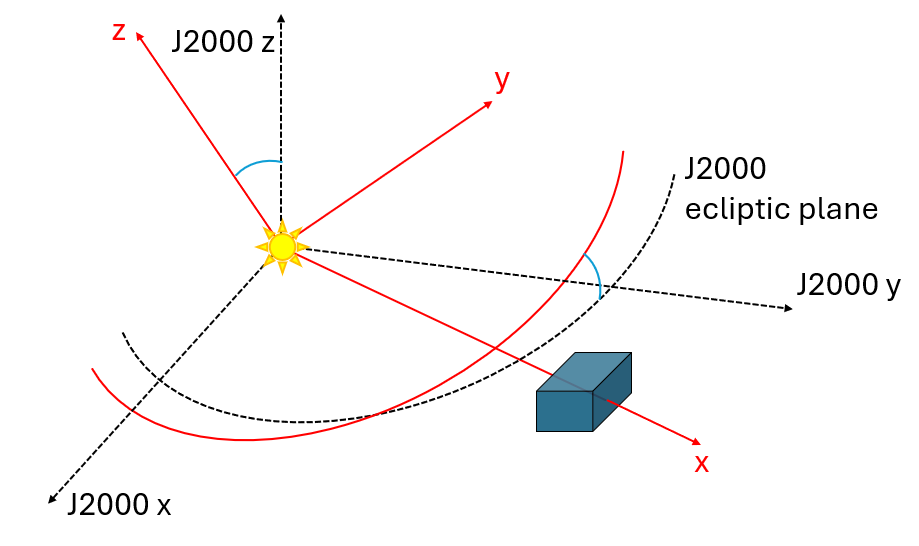}
    \caption{Schematics describing the new reference frame. The solid red lines represent the new reference frame, while the dashed black lines represent the J2000 ecliptic reference frame. The box represents BepiColombo. The two blue angles are of the same value.}
    \label{fig:newframe}
\end{figure}

To assess the damage the particles can cause, we plotted the velocity of the particles in a new reference frame, which we call the BepiColombo reference frame. The spacecraft is at the centre of this new frame, with the x-axis indicating the direction from the Sun to the position of BepiColombo. The z-axis is the rotation of the J2000 z-axis around the J2000 x-axis by the angle between the ecliptic and the newly defined x-axis. This new z-axis is perpendicular to the new x-axis, even if the x-axis is out of the ecliptic plane. Finally, the y-axis completes the frame (see Fig.~\ref{fig:newframe}). This new reference frame is very useful to assess whether the particles come from the Sun or not and from the North of the ecliptic or not.

\begin{figure}
   \centering
    \includegraphics[scale = 0.5]{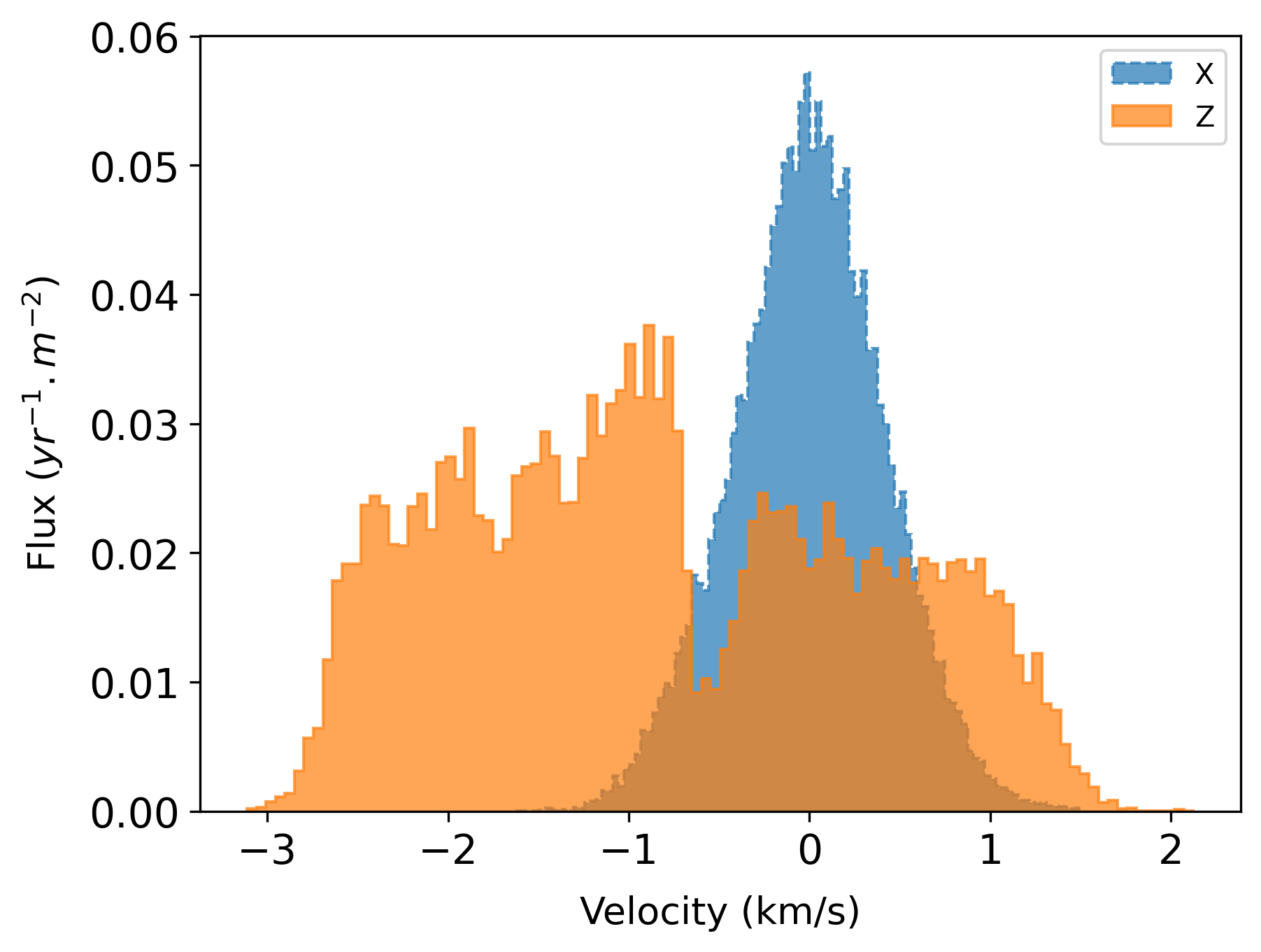}
    \includegraphics[scale = 0.5]{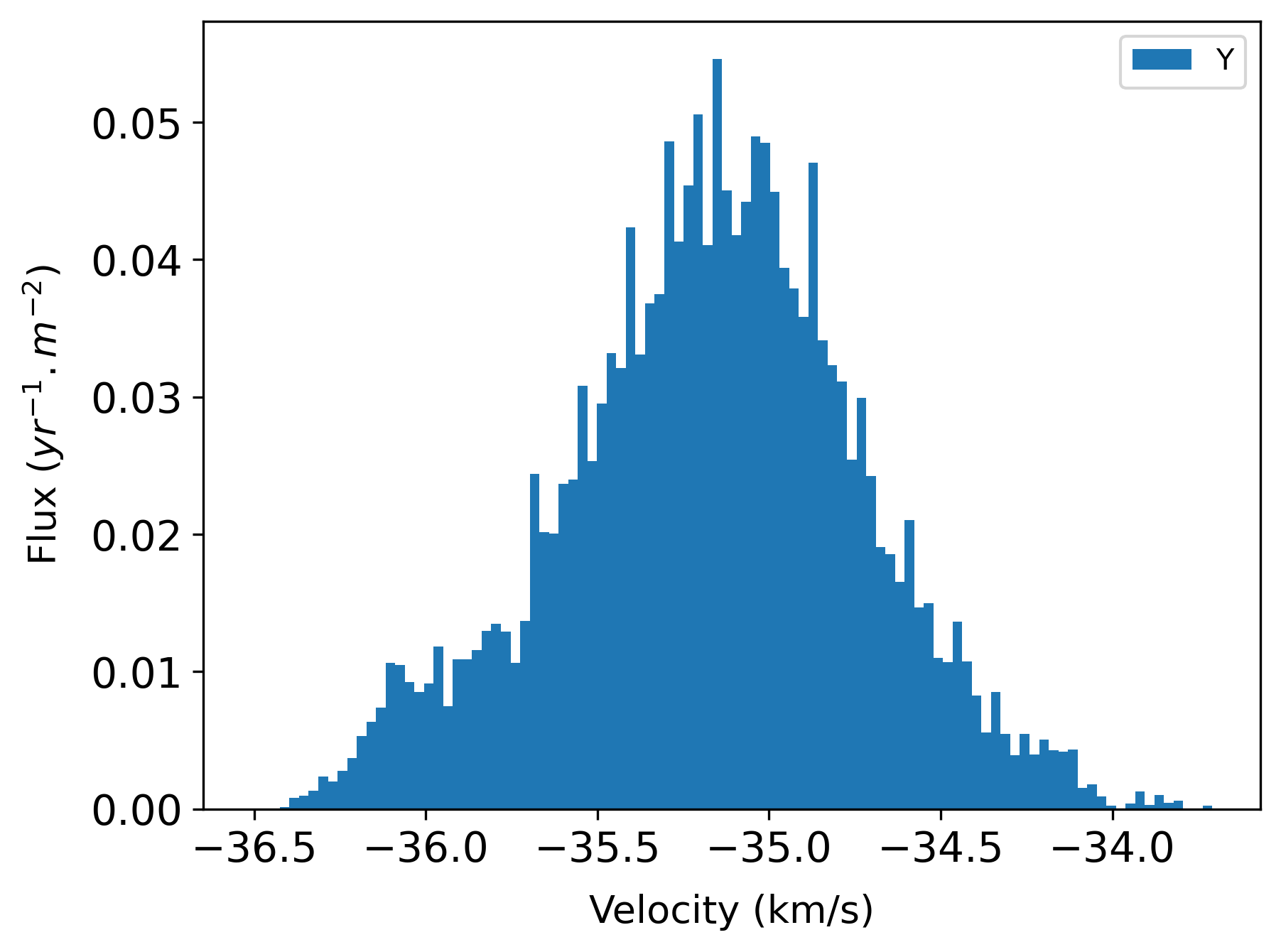}
    \caption{Impact flux as a function of velocity of impacting particles for the three components in the new reference frame}
    \label{fig:speed_newframe}
\end{figure}

Figure~\ref{fig:speed_newframe} show the impact flux as a function of the velocity vector components of the impacting particles in this new frame. This time, this velocity is not computed relative to BepiColombo, as we only look at the direction of the velocity, not the value of its norm. If we suppose that the solar panels are facing the Sun, there are as many particles hitting the front and the back of the solar panels. There are even more particles that are hitting the solar panels on its edges ($V_X = 0$). There are more impacts on the top of BepiColombo, even though impacts on the bottom also happen. Finally, all impacting particles have a negative Y-component. We verified that BepiColombo velocity is also following -Y, which makes sense because BepiColombo is not following a retrograde orbit.

As we have seen, the speed of the particles is not cause for concern, especially since the particles most likely to impact are very small. As for the direction, no particular alarms can be raised following these results since no sensitive parts of the spacecraft are particularly hit by the impact flux.

\section{Discussion}\label{sec:discussion}

This paper focuses on finding all information needed to assess the impact risk a spacecraft like BepiColombo would face when crossing the dust ring. To do this, we had to make some assumptions, which we discuss here, keeping in mind that a meaningful discussion on the impact risks should consider the size distribution, velocity distribution, and the distribution of the direction of impact for each impacting particle, which means all of theses have to be computed.

We thus have to find the position and speed of particles inside the ring, via a model focusing on viable orbits in the ring. By choosing the widest range of orbital elements possible and then selecting orbits that remain in the ring at the end of our 6000~years integration, we are confident that the orbits used for the impact risk computation are representative of the dust ring. Thus the figures shown in Sect.~\ref{sec:res} represent the full evolution of our model and not only of particles that are consistently part of the dust ring.

The length of the integration can also be discussed. A full integration of the ring should last much longer than the 6000~years chosen here. However, given the amount of particles in the ring and the timescales of this project, a longer integration is not possible. More importantly, the purpose of this study is to find the orbits that would best describe the ring, instead of analysing the stability of the ring. There are many ways such a study could be run, with a frequency map analysis such as in \citet{Scholl_al_2005} or with a chaos map, as was done for example in \citet{Courtot_al_2023}. Another approach would be to perform longer term simulations, as in \citet{Cuk_al_2012}. None of the papers cited here performed an analysis of the Venus dust ring itself, but the methods they use on other small particles could be applied for this ring. In the case of the long-term simulation approach, \citet{Pokorny_Kuchner_2019} already used it to study the stability of the ring. They showed that particles captured in the ring must have a diameter greater than 60~$\mu$m and be created in the ring itself. Smaller particles would be pushed out of the ring by solar radiation pressure and particles created outside would be deviated by Poynting-Robertson towards Venus and then pushed outside the ring. We do not seek to confirm these results or to further this kind of study, which is why we do not need to model the evolution of the ring on longer timescales.

Another point to discuss is the use of IMEM2 to obtain the frequency distribution of the masses and sizes of particles in the ring. Using this tool means we are making the hypothesis that the ring follows the same size distribution as the interplanetary dust cloud. This hypothesis would be easily justifiable if the Venus dust ring was created the same way as the Earth dust ring, where particles from the interplanetary dust ring get trapped in mean-motion resonances with the Earth \citep{Reach_2010}. However this hypothesis was disproved by \citet{Sommer_al_2020}. \citet{Pokorny_Kuchner_2019} propose another explanation: they claim that the dust ring is composed in majority of particles from hypothetical co-orbital asteroids and also some particles from Jupiter-family comets. However, strong hypothesis would be necessary to derive from this a reliable frequency distribution of masses or sizes for the particles in the ring. Furthermore, since the co-orbital asteroids have yet to be discovered \citep{Pokorny_al_2020}, there is no information on the composition of the asteroids themselves, which would help make an informed hypothesis on the mass and size frequency distribution of the particles ejected. This discussion shows that using IMEM2 and thus assuming a distribution close to the interplanetary dust cloud is the best available choice for our purpose, which is evaluating the impact risk.

IMEM2 itself has some drawbacks. The model makes several assumptions: the collisions, for example, are treated as sink rather than fragmentations in order to avoid generating particles during integrations. This means the particles are simply removed when they reach the end of their collisional lifetime. The objects from the Kuiper belt are also not included. Finally, gravitational focusing and shielding by planets are not taken into account. Despite these small shortcomings, IMEM2 is used routinely at ESA without any issue.

A final aspect of the IMEM2 results can be discussed: the extrapolation from IMEM2 to obtain impact fluxes on BepiColombo for the smallest particles. Indeed, this extrapolation assumes that such small particles would stay inside the ring as would the bigger ones. However \citet{Pokorny_Kuchner_2019} showed that particles smaller than 60~$\mu$m are not dynamically stable and blend in the zodiacal cloud very quickly. This means flux results are less reliable for this size range.

There is clearly a lack of data about the dust ring. We hope that future spacecraft with a trajectory that cross that region of space can use their instruments to discover more about the ring. Ideally, a dust detector could be installed on a spacecraft crossing the ring, but observations of the ring with a different geometry than the spacecraft mentioned in this paper (Helios, STEREO, the Parker Solar Probe) would also be invaluable, as they would help constrain the ring dimension. Any research on the source of the ring would also be welcomed, in particular concerning the hypothetical co-orbital asteroids that might be responsible for the ring. If these could be found and more information about them were discovered, it would be easier to construct a new and more reliable model of the Venus dust ring.

This model, as we explained throughout this paper, does not pretend to be able to weight on the stability of the ring or its origin. However it does provide a good basis for the impact risk assessment, which was the point of this paper. The description of the particles that we used to compute the number of impacts, could be used to be applied on other spacecraft than BepiColombo and the methodology described in the last section is replicable by anyone, with any dust model that gives the dust flux at a location, similarly to IMEM2.

\section{Conclusion}\label{sec:ccl}
 We have described how we generated initial conditions for particles in the dust ring. The integration of these particles for 6000~years revealed interesting features of the dust ring evolution. The density of particles seems to be higher at Venus's orbit. Close encounters with Venus and NGFs tend to make the dust ring spread out, especially radially. The distribution of the smallest particles shows a peculiar feature in mean anomaly that we could not explain. The reason for this feature is out of the scope of this paper, but future studies could investigate it. 

 By selecting particles that are still in the ring after 6000~years of integration and multiplying them, we obtained reliable conceptualisations of the orbits of the particles. We used these to compute the speed and direction of impacts on BepiColombo, a spacecraft chosen to serve as an example. Both the speeds and directions of the particles do not seem threatening. We also computed the impact flux on BepiColombo using the trajectory of the spacecraft and the model IMEM2, which describes interplanetary dust. The flux is also at a reasonable level and does not seem overly dangerous to a spacecraft following BepiColombo's trajectory.

 Future studies could redo these computations using other models of size distribution in the ring. While IMEM2 is a reliable model, comparison with other models would be welcomed. Such analysis could also be performed for other spacecraft. Indeed, our risk analysis depends on the trajectory of BepiColombo, which only crosses the ring a few times and never stays close to Venus's orbit for long. It is highly possible that a spacecraft orbiting Venus or staying longer in the ring could encounter higher levels of impact flux.

 The biggest drawback of this study is the lack of data on the ring. We have tried to select values that most papers describing the ring agree upon, but this task was very difficult to fulfil. Future missions that point their instruments in the direction of the ring should be able to provide more data that could be used to refine this model and update the risk assessment. Any crossing of the ring could also be an interesting opportunity to study it further.

\begin{acknowledgements}
We warmly thank the referee, Petr Pokorny, for his help in improving this paper.
\end{acknowledgements}

\bibliographystyle{aa}
\bibliography{biblio.bib}

\end{document}